\documentclass[%
twocolumn,
 amsmath,amssymb,
 aps,
 prd,
]{revtex4-2}

\usepackage{graphicx}% Include figure files
\graphicspath{ {plots/} }
\usepackage{dcolumn}% Align table columns on decimal point
\usepackage{bm}% bold math
\newcommand{\snowglobes}{{\sf{SNOwGLoBES}}}

\begin{document}

%\preprint{APS/123-QED}

\title{Triangulation Pointing to Core-Collapse Supernovae with Next-Generation Neutrino Detectors}% Force line breaks with \\
%\thanks{A footnote to the article title}%

\author{N.B.~Linzer}
\email{nlinzer@caltech.edu}
 \altaffiliation[Also at ]{Division of Physics, Mathematics, and Astronomy, California Institute of Technology,
Pasadena, CA 91125 USA}%Lines break automatically or can be forced with \\
\author{K.~Scholberg}%
 \email{schol@phy.duke.edu}
\affiliation{%
 Department of Physics, Duke University, Durham, NC 27708 USA
}%

\date{\today}% It is always \today, today,
             %  but any date may be explicitly specified

\begin{abstract}
A core-collapse supernova releases the vast majority of the gravitational binding energy of its compact remnant in the form of neutrinos over an interval of a few tens of seconds. In the event of a core-collapse supernova within our galaxy, multiple current and future neutrino detectors would see a large burst in activity.  Neutrinos escape a supernova hours before light does, so any prompt information about the supernova's direction that can be inferred via the neutrino signal will help to enable early electromagnetic observations of the supernova.  While there are methods to determine the direction via intrinsic directionality of some neutrino-matter interaction channels, a complementary method which will reach maturity with the next generation of large neutrino detectors is the use of relative neutrino arrival times at different detectors around the globe.  To evaluate this triangulation method for realistic detector configurations of the next few decades, we generate random supernova neutrino signals with realistic detector assumptions, and quantify the error in expected time delay between detections.   We investigate a practical and robust method of estimating the time differences between burst detections, also correcting for detection efficiency bias.   
With this method, we determine the pointing precision of supernova neutrino triangulation as a function of supernova distance and location, detectors used,  detector background level and neutrino mass ordering assumption. Under favorable conditions, the 1$\sigma$ supernova search area from triangulation could be reduced to a few percent of the sky.  It should be possible to implement this method with low latency under realistic conditions.
\end{abstract}

% PACS, the Physics and Astronomy
                             % Classification Scheme.
%\keywords{Suggested keywords}%Use showkeys class option if keyword
                              %display desired
\maketitle

%\tableofcontents

\section{\label{sec:intro}Introduction}
After a massive star has expended all of its fuel, it collapses and can form a compact remnant such as a black hole or neutron star. Such a core-collapse supernova releases a huge burst of neutrinos.  These neutrinos escape the star within a few tens of seconds of the collapse, and have energies on the order of a few to several tens of MeV~\cite{Mirizzi2015}.  The burst of neutrinos can be detected on Earth \cite{Scholberg2012}.  The few dozen neutrino events observed from SN1987A in the Large Magellanic Cloud comprise the first and only such detection~\cite{Bionta:1987qt,Hirata:1987hu,Alekseev:1987ej,Aglietta:1987it, Vissani:2014doa}.   The current generation of detectors has capability for detection of a few orders of magnitude more events, and the next generation will have yet another order of magnitude in reach, as well as richer flavor sensitivity~\cite{Scholberg2012}.
 
Current-generation detectors with the sensitivity to detect the burst of neutrinos associated with supernovae include Super-Kamiokande (Super-K)~\cite{Abe:2016waf}, LVD~\cite{Agafonova:2014leu}, Borexino~\cite{Monzani:2006jg}, KamLAND~\cite{Abe:2008aa}, and IceCube~\cite{Halzen:1994xe, Halzen:1995ex,Abbasi:2011ss,Kopke:2017req}, as well as HALO~\cite{Duba:2008zz,Vaananen:2011bf}, Daya Bay~\cite{Wei:2015qga} and NOvA~\cite{Vasel:2017egd}. Detectors coming online in approximately the next decade include Hyper-Kamiokande (Hyper-K)~\cite{Abe:2011ts}, DUNE~\cite{Acciarri:2015uup}, and JUNO~\cite{An:2015jdp}. These detectors have different detection mechanisms and flavor sensitivities. Water-based Super-K and the planned Hyper-K employ Cherenkov radiation.   Liquid scintillation detectors, in contrast, monitor scintillating compounds in liquid organic hydrocarbons that release photons in response to charged particles; examples of this type of detector include LVD, Borexino, and JUNO. Other detectors, such as DUNE and HALO, employ liquid argon or lead, respectively~\cite{Scholberg2012, Mirizzi2015}.  Both water and scintillator detectors are sensitive primarily to the $\bar{\nu}_e$ component of the supernova flux, via inverse beta decay on free protons.  Argon will have excellent sensitivity to the $\nu_e$ component of the flux, via charged-current neutrino absorption on argon nuclei.

 Compared to the rapid escape of weakly-interacting neutrinos during star collapse, photons emerge more slowly.  This enables an early alert for a core-collapse supernova~\cite{Antonioli:2004zb,Scholberg:2008fa}.  The additional value of prompt pointing information from the neutrino signal is obvious:  without directional information, astronomers will not know where to look for the supernova. Some core collapses may not lead to bright and obvious events in electromagnetic radiation, due to collapse to a black hole or other explosion failure~\cite{OConnor:2010moj}.   Such failed supernovae may still be quite bright in neutrinos, and in these cases,  pointing information is still valuable as it could help to narrow the search for a ``winked-out" progenitor~\cite{Kochanek:2008mp}.   Furthermore, knowing the direction of the supernova signal, even in the absence of an identified supernova or a progenitor, will aid in evaluating the neutrino trajectory in the Earth in order to estimate neutrino matter effects~\cite{Lunardini:2001pb,Lunardini:2003eh,Dighe:1999bi,Mirizzi2015}.
 
 The neutrino burst signal in an individual detector can be used for pointing.  This requires both neutrino-matter interactions with intrinsic directionality, and detector technology with the capability to exploit such directionality. Other publications have explored supernova pointing with neutrinos~\cite{Burrows:1991kf,Beacom1999,Tomas:2003xn,Scholberg:2009jr,Muhlbeier2013,Fischer:2015oma,Brdar2018,Hansen:2019giq}.  There are excellent prospects for use of neutrino-electron scattering in Super-K, which should yields few-degree pointing~\cite{Abe:2016waf}.  Pointing using fine-grained tracking in DUNE~\cite{ajroeth} is also very promising.  Other possible methods involve other anisotropic interactions, as well as Earth-matter-oscillation-based pointing~\cite{Scholberg:2009jr}.
Triangulation pointing methods have been explored in several of the references mentioned above~\cite{Burrows:1991kf,Beacom1999,Muhlbeier2013,Brdar2018,Hansen:2019giq}. 

Even given the existence of effective single-detector pointing methods based on anisotropic interactions, the best approach is to exploit all possible prompt information in the context of multi-messenger astronomy~\cite{GBM:2017lvd}.  Latency also matters for real-time astronomy. The more rapid the pointing, the more likely it will be to locate early supernova light.

In this paper we revisit the triangulation question using realistic detector response assumptions using the \snowglobes ~\cite{snowglobes}~event rate calculator.  We take a somewhat different, and practical, approach with respect to those in Refs.~\cite{Brdar2018,Hansen:2019giq}. Rather than evaluating time resolution for a given detector with respect to the ``true" neutrino wavefront time, we instead focus on evaluating the variance of burst time difference determination between detectors, and apply a correction for bias due to difference in detector response.  We consider a robust corrected-first-event method which should be relatively straightforward to implement in practice.  Using this method, we also consider pointing precision as a function of distance and location on the sky, as well as mass ordering assumption.

Section~\ref{sec:method} describes our methods for calculating event rates,  estimating the variance of burst timing differences between detectors, and estimating the pointing precision in terms of constrained area on the sky for a given detector configuration assumption.   Section~\ref{sec:results} presents selected results as a function of distance to the supernova, detector combination, supernova model, and mass ordering. Conclusions are provided in Sec.~\ref{sec:conclusion}.

\section{Methods}\label{sec:method}

\subsection{Neutrino Event Rate Calculations}
The \snowglobes~event rate calculator folds fluxes, cross-sections, and detector smearing to determine mean expected neutrino interaction signals in multiple current and future detectors.  To describe an expected neutrino signal as a function of time over the $\sim$10~seconds of a burst, we assume a ``pinched-thermal"  spectrum parameterized by the following functional form 
(e.g.,~\cite{Minakata:2008nc,Tamborra:2012ac}):
\begin{equation}
        \label{eq:pinched}
        \phi(E_{\nu}) = \mathcal{N} 
        \left(\frac{E_{\nu}}{\langle E_{\nu} \rangle}\right)^{\alpha} \exp\left[-\left(\alpha + 1\right)\frac{E_{\nu}}{\langle E_{\nu} \rangle}\right] \ ,
\end{equation}
where $E_{\nu}$ is the neutrino energy, $\langle E_\nu \rangle$ is the
mean neutrino energy, $\alpha$ is a ``pinching parameter'', and
$\mathcal{N}$ is a normalization constant related to the energy release.   The parameters are specified for each observable flavor, $\nu_e$, $\bar{\nu}_e$ and $\nu_x$ (where $\nu_x$ represent the sum of $\nu_{\mu}$, $\bar{\nu}_\mu$, $\nu_\tau$ and $\bar{\nu}_\tau$) as a function of time to describe the neutrino energy and flavor evolution over the burst.  This treatment is fairly standard in the literature.  \snowglobes~ computes event rates for any specified time bin.

% Add pinching parameter eqn
For our baseline treatment of the time-dependent neutrino signal we used the pinching parameters as a function of time from an electron-capture 8.8-M\textsubscript{\(\odot\)} supernova~\cite{Hudepohl2010} (the so-called ``Garching model") to describe the spectrum as a function of time.  This is a relatively low-neutrino-flux model. From these parameters, we generate neutrino fluxes as a function of time and energy. \snowglobes ~subsequently folds the fluxes with cross sections of detector materials, such as water, argon, or liquid scintillator, and assumed detector responses (efficiencies and smearing).  The dominant event channel for water and scintillator detectors is inverse beta decay, $\bar{\nu}_e + p \rightarrow e^+ + n$, and for argon the dominant channel is $\nu_e + {}^{40}{\rm Ar} \rightarrow e^- + {}^{40}{\rm K}^*$; however in all detector cases, there are subdominant contributions from all flavors~\cite{Scholberg2012,Mirizzi2015}.
The default cross-section and detector-response assumptions for \snowglobes~version 1.2 are used, and all relevant neutrino interaction channels, including subdominant channels, are included in the event count.   We then generate the expected neutrino signals as seen in multiple different detectors. For these simulated supernova signals, we generate the mean expected number of neutrinos detected within 0.5-ms bins, with a total signal time window of 10~seconds. Supernova event rates are computed at a nominal distance of 10~kpc.  Event rates are scaled by the inverse square of distance to find rates at different distances.  Table~\ref{tab:det_assumptions} summarizes the assumptions used.

For these studies, we assume that the resolution on the absolute  arrival time for individual neutrino events is much less than the 0.5-ms bin size.  In general, absolute event-time stamping at the tens of ns level is feasible using GPS; this is well validated by long-baseline neutrino experiments which use accurate and precise GPS timing to identify beam events. Scintillator and water experiments, which record photons, have intrinsically precise timing at better than the tens of ns level.  Liquid argon time projection chambers like DUNE, which drift ionization signals over several ms,  must use the scintillation photon component of their signal to achieve individual-event timing at the sub-ms level~\cite{Acciarri:2015uup}.

\begin{table}[h]
 \caption{Detector assumptions used for this study.\footnote{Note that while fiducial masses are used, in practice, expanded volume is a possibility for a burst detection.}  \label{tab:det_assumptions}}
\centering
\begin{tabular}{|c|c|c|c|}

\hline
Detector & Material & Mass (kton) & Detector Location\\ \hline \hline
Super-K & Water & 22.5 & 36.2$^{\circ}$ N, 137.2$^{\circ}$ E\\
Hyper-K & Water & 374 & 36.4$^{\circ}$ N, 137.3$^{\circ}$ E\\
DUNE & Argon& 40 & 44.4$^{\circ}$ N, 103.8$^{\circ}$ W\\
JUNO & Scintillator & 20 & 22.1$^{\circ}$ N, 112.5$^{\circ}$ E\\ \hline

\end{tabular}
\end{table}

\begin{figure}[ht]
\centering
\includegraphics[scale=0.5]{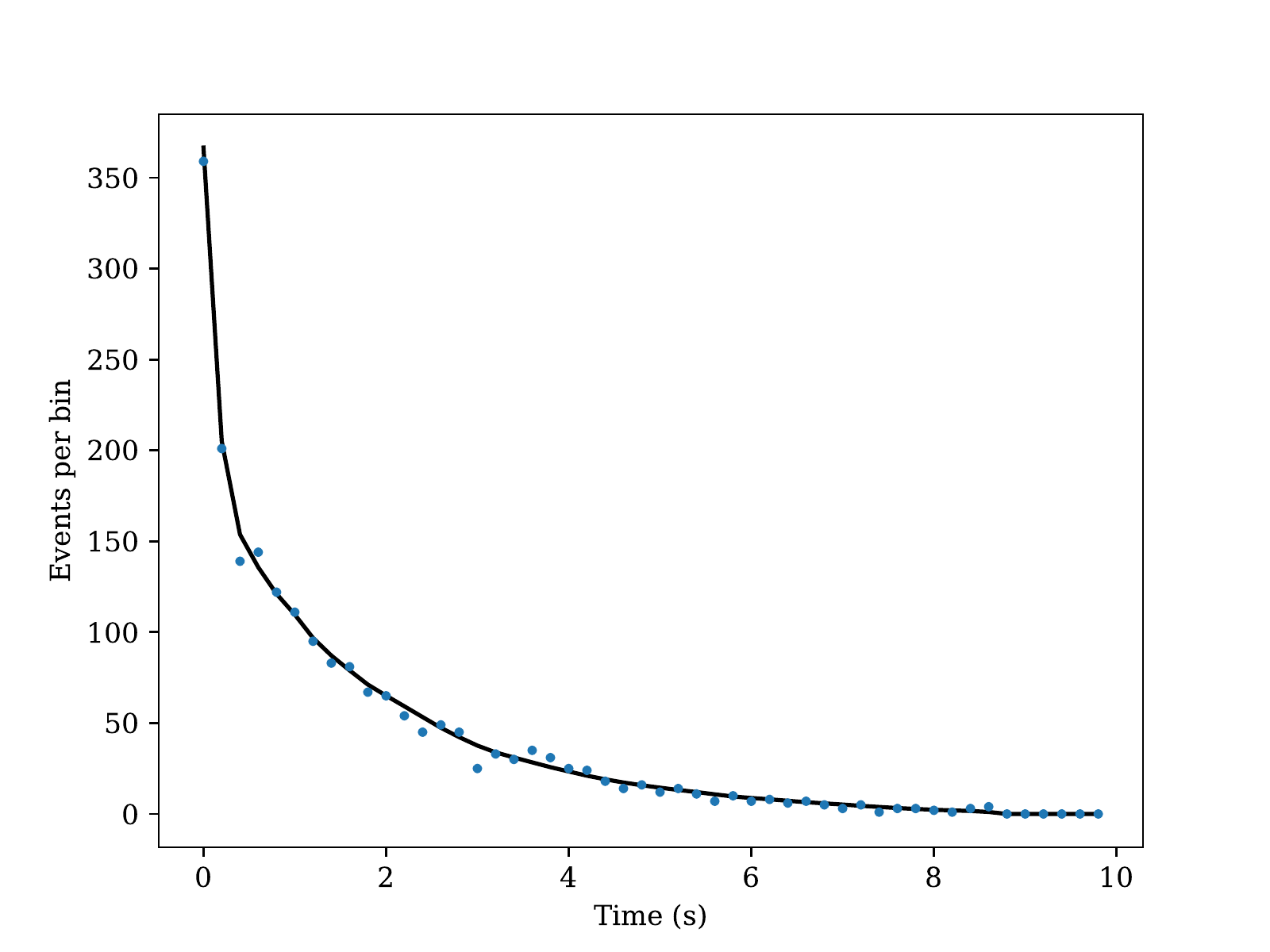}
\includegraphics[scale=0.5]{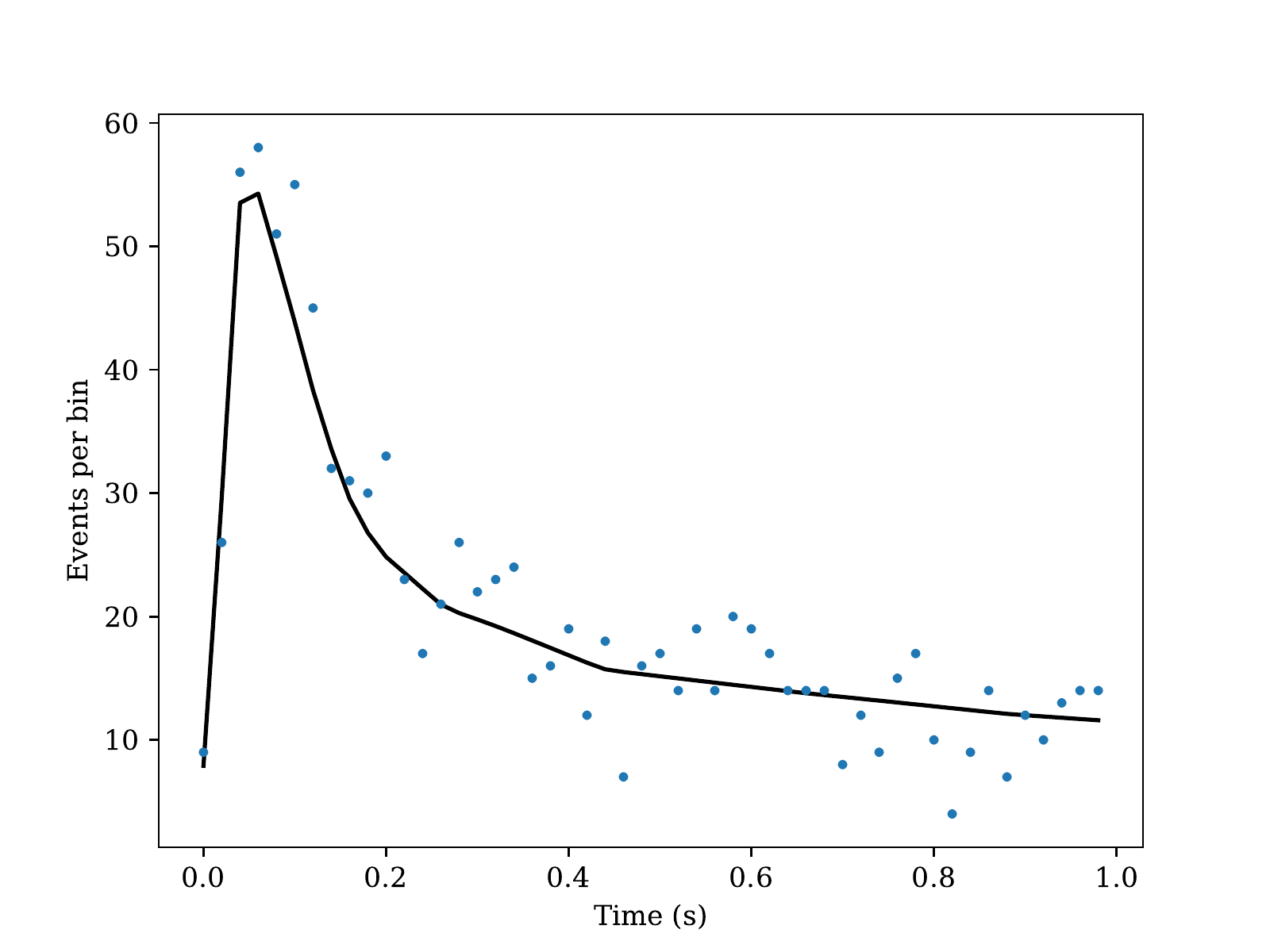}
\caption{The solid line gives the expected neutrino event rate in Super-K in events per time bin as generated by \snowglobes. The points show one instance of a randomly-generated supernova signal at 10~kpc. Both plots have larger binning than that used in the analysis. In the top plot, the bins are 200~ms wide for a full signal time of 10~s. The bottom plot has 20-ms bins for a total time of 1~s. This example assumes no flavor transitions. }
\label{fig:sk_rand}
\end{figure}

Using the expected mean event rates generated by \snowglobes, we simulate random supernova events by fluctuating the contents of each time bin according to a Poisson distribution with mean set to the expected value calculated by \snowglobes. The total numbers of events for each detector are in Table~\ref{tab:event_rates}.  This creates a random neutrino signal spectrum for a given detector. An example of a simulated time-dependent signal for Super-K is given in Fig.~\ref{fig:sk_rand}.

\begin{table}[h]
 \caption{Expected numbers of events in different detectors from \snowglobes ~for the Garching model~\cite{Hudepohl2010}, for the first 10 seconds and the first 10 ms for a 10-kpc supernova.  The columns labeled IO give the event counts under the inverted ordering assumption, and the columns labeled NO give the event counts for the normal ordering assumption (see Sec.~\ref{sec:ordering}). \label{tab:event_rates}}
\centering
\begin{tabular}{|c|c|c|c|c|}

\hline
Experiment & IO, 10 s & IO, 10 ms & NO, 10 s & NO, 10 ms\\ \hline \hline
Super-K & 2170 & 1.8 & 2110 & 0.6\\
Hyper-K & 36076 & 29 & 35081 & 11\\
DUNE & 1414 & 34 & 1568 & 1.2\\
JUNO & 2868 & 4.8 & 2798 & 2.3\\ \hline

\end{tabular}
\end{table}

\subsubsection{Mass-Ordering-Dependent Flavor Transition Effects}\label{sec:ordering}

The supernova neutrino signal is also affected by neutrino flavor transitions, which in general depend on the neutrino mass-state ordering, or hierarchy~\cite{Mirizzi2015}.  In the standard three-flavor neutrino paradigm, there are three neutrino masses, $m_1$, $m_2$, and $m_3$.  We have information from multiple oscillation experiments on two mass-squared difference scales, one $\sim 7.5 \times 10^{-5}$eV$^2$, and the other $\sim 2.5\times 10^{-3}$ ~eV$^2$~\cite{Esteban:2016qun}.    The world data are consistent with two light and one heavy mass states, a situation referred to as the normal ordering (NO).  Also consistent with the data are one light and two heavier mass states, which is referred to as the inverted ordering (IO).   It is likely that the mass ordering will be known with reasonable significance from long-baseline and/or reactor neutrino oscillation experiments within the next decade~\cite{Patterson:2015xja}.

Neutrino flavor transitions (often referred to as ``oscillations"\footnote{Note this nomenclature is not always strictly applicable~\cite{Smirnov:2016xzf}}) modulate the time, energy and flavor structure of the supernova neutrino burst~\cite{Scholberg2017}. Within the supernova itself, these flavor transitions can be due to matter  (Mikheyev-Smirnov-Wolfenstein, MSW) effects, or due to self-induced flavor effects (neutrino-neutrino interactions), sometimes known as ``collective effects."  In both cases, the specific modulation depends on the mass ordering.  Either effect can dominate depending on whether the matter potential or neutrino-neutrino potential is dominant. 
In the MSW case, assuming adiabatic transitions, the flavor modulation can be treated relatively simply, according to:
$ F_{\nu_e} =F^0_{\nu_x}$ and 
 $F_{\bar\nu_e} = \cos^2 \theta_{12} F^0_{\bar\nu_e} + \sin^2 \theta_{12} F^0_{\bar\nu_x}$ for NO, and
 $F_{\nu_e} =  \sin^2 \theta_{12} F^0_{\nu_e} + \cos^2 \theta_{12} F^0_{\nu_x}$, and
 $F_{\bar\nu_e} =  F^0_{\bar\nu_x}$  for IO, 
where $F(\nu_i)$ is the flux of a given flavor ($F(\nu_x)$ represents the flux of any of either $\nu_\mu$ or $\nu_\tau$, and similarly for antineutrinos), and $\theta_{12}$ is the relevant ``solar" mixing angle.  Matter-induced flavor transitions also occur when neutrinos traverse the Earth, but this is a percent-level or smaller effect and we ignore it here~\cite{Mirizzi2015}.

\begin{figure}[ht]
\centering
\includegraphics[scale=0.5]{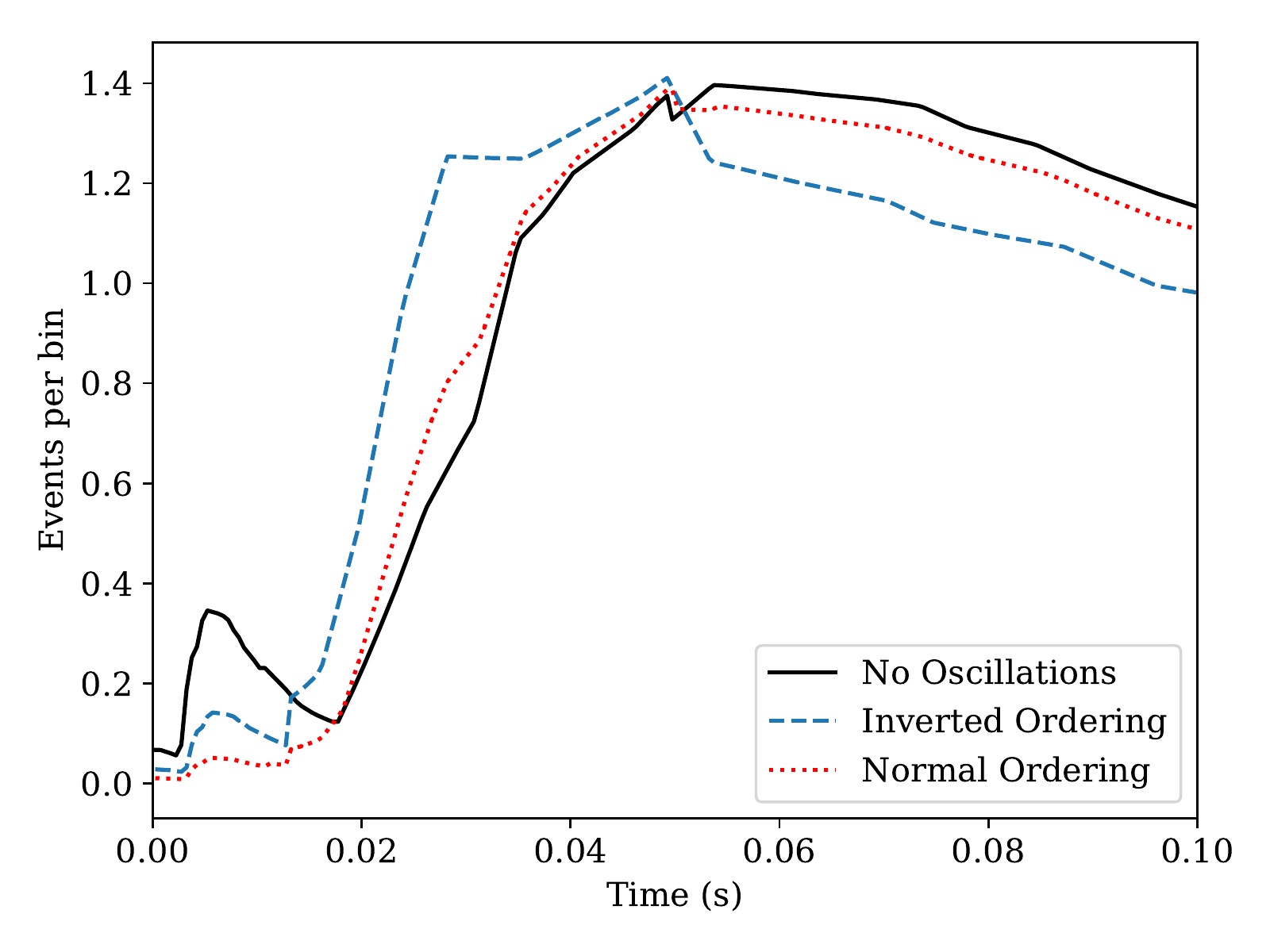}
\caption{Expected neutrino signal observed at Super-K for different mass ordering assumptions with 0.5-ms bins. When mass ordering assumptions are added, the event rate at the beginning of the signal is reduced, especially in the case of normal ordering.}
\label{fig:osc}
\end{figure}

It is expected that the dominant flavor transition effect at early times within the burst will be the standard MSW effect described by the equations above~\cite{Mirizzi2015}.  This effect can create a strong modulation of the neutronization burst within a few tens of milliseconds of the start of the signal.  Figure~\ref{fig:osc} demonstrates this effect for the model from Ref.~\cite{Hudepohl2010}.  The neutronization burst, originally $\nu_e$,  is strongly suppressed in the NO case, and moderately suppressed in the IO case. The specific effect will depend on the turn-on profile of the other flavors. 

This flavor transition effect on the time profile strongly affects the accuracy of the first-event method, which increases with higher event rate at the beginning of the signal.  Because flavor transitions tend to reduce the event rate at early times, they tend to reduce the precision of this triangulation method.

\subsection{Estimating the Time Difference Between Neutrino Pulses}

Triangulating a supernova requires estimating the time difference between neutrino pulse sampled at multiple detectors.  We used a Monte Carlo to test several methods for finding this time difference using the simulated samples generated using the methods described above.
For a given trial, we generate two random detector time-profile signals, and offset one detection from the other by a fixed time shift. Approximating the neutrinos' speed as the speed of light, the maximum time difference between two detected signals on Earth would be $\sim$40 ms. For these analyses, the applied offset was 4 ms. However, the results do not depend on the value of the assumed time shift. For each detector combination among DUNE, JUNO, Super-K, and Hyper-K, we generate 10,000 pairs of random burst signals. Then, we find the burst time difference using a given time-difference-finding method. 
The variance of the distribution of these time differences allows us to estimate the expected time difference error, $\sigma_t$, as its square root.

We tried a chi-squared best-fit method, a cross-correlation method, and a method based on finding the time of maximum event rate,  but found that a method based on simple comparison of the first event times in the burst had the lowest variance.  Furthermore, given that such a method would in practice be straightforward to implement and would require minimal prompt data analysis for sharing among experimental collaborations, it is likely to be both robust and practical for low latency.  We therefore focus on this method in the remaining studies described here.

We note that the burst time can be found more accurately by fitting to the expected event rate calculated by \snowglobes.  However, in the case of a real detection we would not know the functional form of the underlying true event rate, given uncertainty in the expected signal.  Not only are there uncertainties in the model assumptions, but there may also be considerable variations due the nature of the progenitor~\cite{OConnor:2012bsj, Hudepohl2013}. Therefore, any approach which relies on comparison of data to a specific model may not be robust, and estimates based on fits to a specific assumed rate, or even an assumed functional form, such as in Ref.~\cite{Brdar2018}, may result in overoptimistic results. We therefore require our triangulation method to make use only of time differences between computed observations.

\subsubsection{Effect of Backgrounds}
\begin{figure}[ht]
\centering
\includegraphics[scale=0.5]{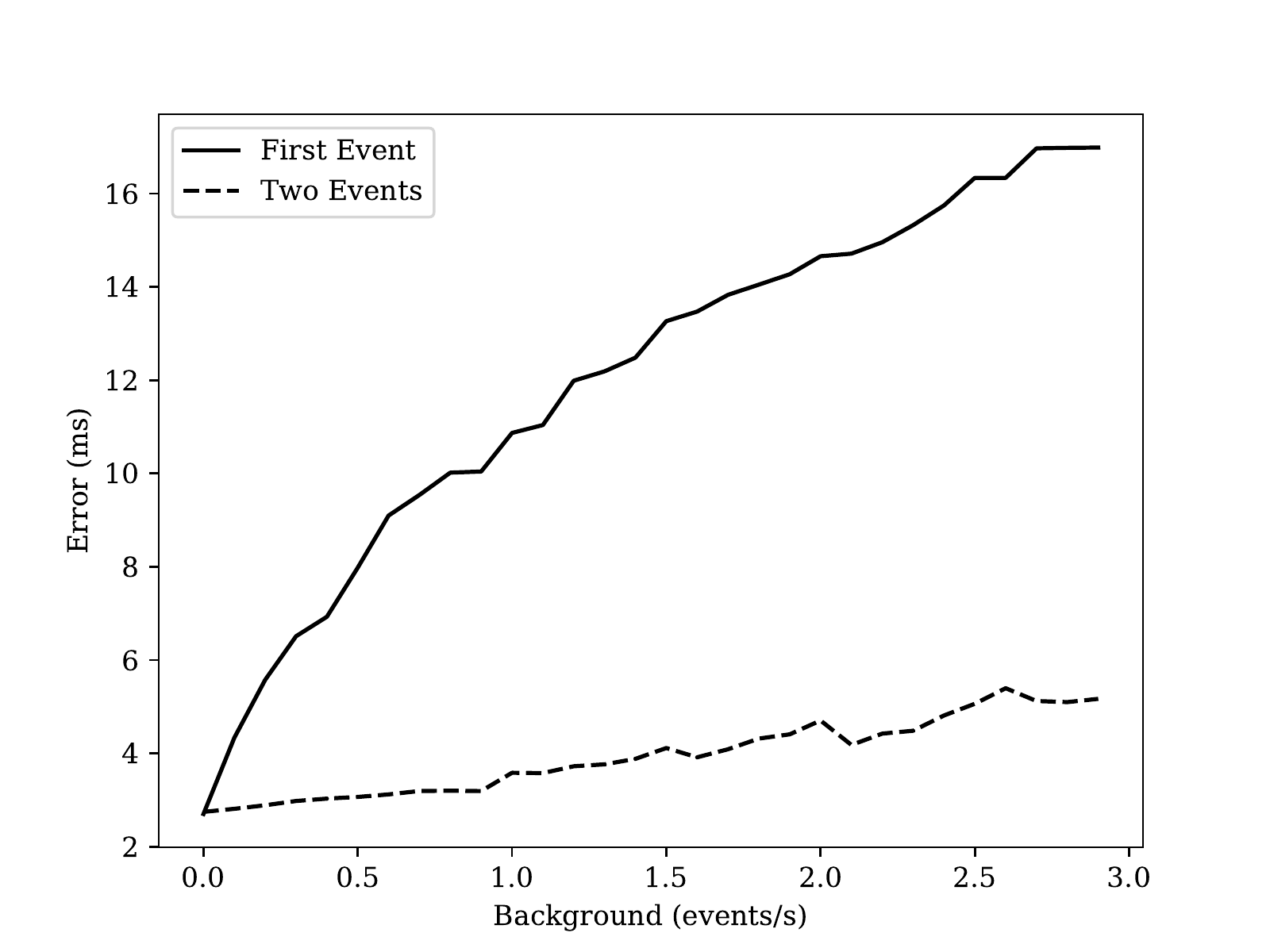}
\caption{Standard deviation of calculated time difference between Super-K and JUNO as a function of background rate. By using the first event that is followed by an additional event within 15~ms, the error is greatly reduced compared to only using the first event. This error only includes the variance in time shift, not any additional error associated with correcting for the bias in the mean.}
\label{fig:sk_juno_bkg_err}
\end{figure}

The arrival time of the first observed event will be dependent on the background rate, as well as the length of time window before the true detection begins. As seen in Fig.~\ref{fig:sk_juno_bkg_err}, as background rate increases, the error in time shift between two detected signals also increases. 

To reduce this background-induced error, we select the first neutrino event for which at least one other event follows soon after. In this analysis, we require at least one other event in a window of 15 ms. This method of first-event selection reduces the $\sigma_t$, as seen in Figure \ref{fig:sk_juno_bkg_err}. The background rate in Super-K is approximately 0.01 Hz~\cite{Abe:2016waf} for an energy threshold of 7~MeV, and error in the first-event method is low at this rate. However, the background rate may be much greater for future detectors.   For the remaining analysis, we conservatively assume a background rate of 0.1 events/s/20~kton.  For all simulated detected signals, a period of 50 ms of only background is added before the start of each simulated signal burst.  In practice, a specific first-event selection could be optimized for known background rates.

\subsubsection{Bias Correction for Detector Response Differences}\label{sec:firstevent}
 
\begin{figure}[ht]
\centering
\includegraphics[scale=0.5]{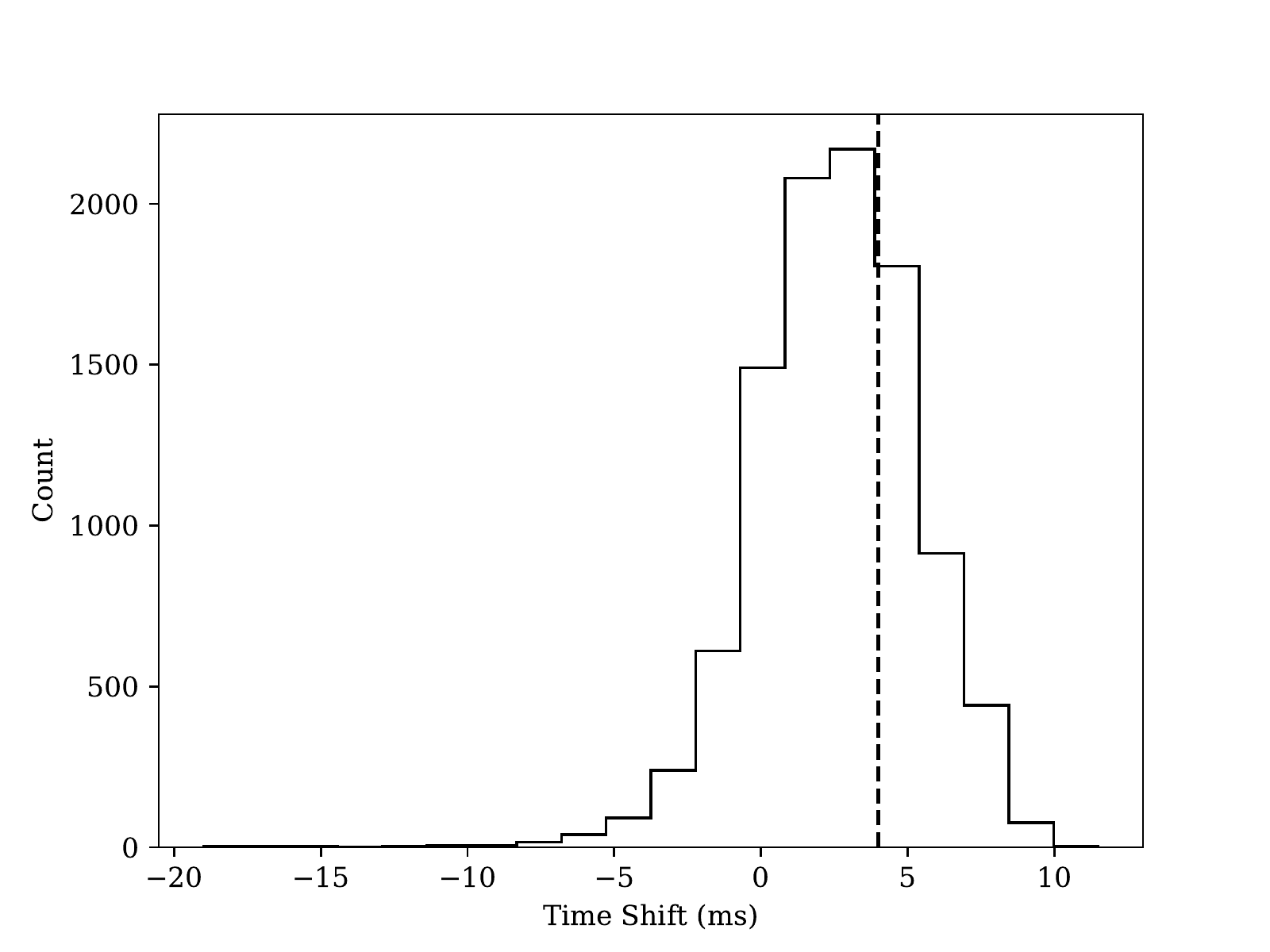}
\caption{Distribution of time differences $\Delta$ calculated by comparing the first event times in two randomly-generated detected signals. In this case, the two simulated detectors are Super-K and JUNO. The dotted line gives the true time difference added between the two signals, 4 ms. The mean of the distribution is 2.5~ms with a standard deviation of 2.7~ms over a total of 10,000 samples.}
\label{fig:sk_juno_no_bkg}
\end{figure}

 In Fig.~\ref{fig:sk_juno_no_bkg}, we show an example of the distribution of time differences when comparing JUNO and Super-K.

An obvious issue with the simple first-event method is that there will be a bias in first event time difference between detectors if the detectors are not identical, due to different expected mean rates near the start of the signals.  The event time profile varies according to detector mass, flavor sensitivity and detection threshold and efficiency. The first event will be more likely to be detected earlier if the mean rate is higher in early time bins.  Therefore, when comparing first events, even if there is low variance in the calculated time difference,  there will be a bias in the mean with respect to the true time offset. This bias is unique to each detector combination, and changes with the specific time profile observed in each detector. We can estimate the bias between two detectors by finding the calculated time difference over many trials and comparing to the actual offset.  Furthermore, in practice, it should be possible to (partially) correct for the bias by making use of the data themselves--- the observed time profile beyond the first event can be used to estimate a correction to the first event time.  

Given an event time profile $R(t)$, the probability of the first event happening at time $t$ is

\begin{equation} 
\label{eq:prob_t0}
P(t) = e^{-\int_{t_0}^{t} R(t') dt'}
\end{equation}

\noindent where $t_0$ is the time the signal begins, and $R(t)$ is the event time probability distribution.
%In the first-event method, $t_0$ is the time of the first event in the randomly generated signal $R(t)$. 
If we have two detected first-event times in detectors $i$ and $j$, the probability of a given time difference, $\Delta = t_i-t_j$, between them is, for $\Delta>0$,

\begin{equation} 
\label{eq:pos_delt}
P_1(\Delta) = \int_{t_{0j}}^{\infty} P_i(t + \Delta) P_j(t) dt
\end{equation}

\noindent where $P_i(t)$ and $P_j(t)$ are given by Eq.~\ref{eq:prob_t0}, for observed $R_i(t)$ and $R_j(t)$ respectively and corresponding to two different observed neutrino signals in detectors $i$ and $j$, and where $t_{0j}$ is the start time for detector $j$. If $\Delta < 0$, we have a similar expression:

\begin{equation} 
\label{eq:neg_delt}
P_2(\Delta) = \int_{t_{0i}}^{\infty} P_i(t) P_j(t - \Delta) dt
\end{equation}

\noindent The expectation value of this time difference for $t_{0i}=t_{0j}$ is the bias $B_{ij}$, and can be found by making use of: 

\begin{equation} 
\label{eq:exp}
B_{ij}= \langle \Delta \rangle = \frac{\int_{0}^{\infty} P_1(\Delta) \Delta d\Delta}{\int_{0}^{\infty} P_1(\Delta) d\Delta} + \frac{\int_{-\infty}^{0} P_2(\Delta) \Delta d\Delta}{\int_{-\infty}^{0} P_2(\Delta) d\Delta}
\end{equation}

\noindent We can use these expressions to 
estimate a bias correction for use when determining the time difference between the first events in two samples. Because the underlying event distribution is not known, we must approximate the distribution using the observed event profiles, which are random samples from $R_i(t)$ and $R_j(t)$.  In the simulations, we take $t_0$ as the time of the first event in the randomly-generated signal in the evaluation of Eq.~\ref{eq:prob_t0}.
This data-estimated bias correction, therefore, does not fully account for the true bias. Also, because information from the detected event profile is being used to find the bias, the error $\sigma_t$ increases in the bias-corrected time distribution. In almost all detector combinations we tested, the bias was reduced using this data-estimated correction, but never by the full amount. In most cases, the bias was reduced by a greater amount than the error increased.  The bias correction improves accuracy but not necessarily precision.  Table~\ref{tab:biases} shows the results for simulated pairs of detectors.

\begin{table}[h]
 \caption{ Estimated timing errors and biases for different detector combinations for a 10-kpc supernova with IO. First column: detector combination.  Second column: timing difference error for that pair.  Third column:  Bias of mean time difference with respect to true time difference.  Fourth column:  Bias estimated from the data according to Eq.~\ref{eq:exp}.  Fifth column: time difference error following bias correction.  Sixth column:  bias after correction.  All quantities in ms. \label{tab:biases}}
\centering
\begin{tabular}{|c|c|c|c|c|c|}

\hline
Detector combination & $\sigma_{t}$ & $B_{{\rm true}}$ & $B_{{\rm est}}$ & $\sigma_{t,~{\rm corr}}$  & $B_{{\rm corr,true}}$\\ \hline \hline
Super-K, DUNE&4.2&5.0&2.5&5.3&2.6\\
Hyper-K, DUNE&1.6&-0.6&0.01&2.0 &-0.6\\
Super-K, JUNO&4.7&3.3&1.4&5.7&1.9\\
Hyper-K, JUNO&2.6&-2.3&-1.0&3.1&-1.3\\
DUNE, JUNO&2.5&-2.3&-1.1&3.0&-1.3\\
 \hline

\end{tabular}
\end{table}

In practice, a bias correction could be done promptly if experimental collaborations shared with each other their $P(t)$ distributions estimated from the observed burst event rates.  If it is not feasible to provide this information in near-real time,  one could consider using estimated corrections based on reasonable models and total event rates.  Optimization of a practical near-real-time triangulation strategy will be the subject of a future investigation.

We note that some of the intrinsic first-event-time bias is due to different flavor sensitivities of different detectors, because there is an expected variation in early flavor content as a function of time (all detectors, however, are sensitive to elastic scattering on electrons, which is likely to dominate the early low-energy signal, and which will minimize this bias).   This effect is taken into account in these \snowglobes ~studies, although we do not attempt to take it explicitly into account in the bias correction.  In principle, a flavor-dependent bias correction could be devised given the detectors' respective flavor-tagging capabilities, if there is sufficient prompt information exchange.

\subsubsection{Including IceCube}\label{sec:icecube}

Long-string detectors in water and ice like IceCube~\cite{Abbasi:2011ss} and KM3NeT~\cite{Molla:2019nns} require special handling for this study.
Such detectors do not observe supernova neutrino signals in the same event-by-event way as do other detectors. Rather, they make use of a single photoelectron count excess over a large dark-rate background.  We take IceCube as an example here and estimate how its data could contribute to the pointing. IceCube's position at the South Pole is especially advantageous in combination with the other detectors for the purpose of triangulation.   Results from fitting the signal start time~\cite{Halzen2009} give the reconstructed start time for a core collapse 10~kpc away as $t_0 = 3.2 \pm 1.0$ ms compared to a true bounce time of t = 0 s when assuming normal mass ordering \cite{Halzen2009}. We use this information for triangulation in combination with the others by finding the first-event-time bias and error for all other detectors individually. Then we can find the net bias and total error to use with Eq.~\ref{eq:chi}. This method does not allow us to apply any bias correction for IceCube, but does provide an estimate for error and bias when combining IceCube information with additional detectors.  The promising results, described in Sec.~\ref{sec:results}, could likely be improved with a dedicated simulation study including IceCube (or KM3NeT).

\subsection{Triangulation Precision}

To estimate the precision of this triangulation method in terms of sky  area, we apply the method described in Ref.~\cite{Brdar2018} with some adjustments. It is assumed that the neutrino signal occurs on the vernal equinox at noon for ease of calculation. Additionally, the location of the supernova is set at the galactic center, at right ascension $\alpha$ = -94.4$^{\circ}$  and declination $\delta$ = -28.92$^{\circ}$. As in Ref.~\cite{Brdar2018}, the time difference between two detectors located at $\textbf{r}_i$ and $\textbf{r}_j$ in the equatorial coordinate system is given by

\begin{equation} 
\label{eq:time}
t_{ij} = \frac{(\textbf{r}_i - \textbf{r}_j)\cdot \textbf{n}}{c}
\end{equation}

\noindent where the neutrino speed is approximated as the speed of light, and \textbf{n} is the direction from which the neutrinos arrive, and is defined as

\begin{equation} 
\label{eq:norm}
\textbf{n} = (-\sin{\alpha} \cos{\delta}, -\sin{\alpha}\cos{\delta}, -\sin{\delta})
\end{equation}

\noindent To find the 1$\sigma$, 2$\sigma$, and 3$\sigma$ areas, we then apply the chi-squared formula defined in Ref.~\cite{Brdar2018}, although including bias correction:

\begin{equation} 
\label{eq:chi}
\chi^2(\alpha, \delta) = \sum_{i, j}^{i < j} \bigg(\frac{(t_{ij}(\alpha', \delta') + B_{ij}) - t_{ij}(\alpha, \delta)}{\sigma_{t,ij}}\bigg)^2
\end{equation}

Here, $t_{ij}(\alpha', \delta')$ is the true time difference between detections given the supernova location, ($t_{ij}(\alpha', \delta') + B_{ij}$) is the expected measured time difference, $\sigma_{t,ij}$ is the calculated time difference error, and $B_{ij}$ is the mean bias estimated by the method described in Sec~\ref{sec:firstevent}.

\section{Results}\label{sec:results}

\begin{figure*}[ht]
\centering
\includegraphics[scale=0.3]{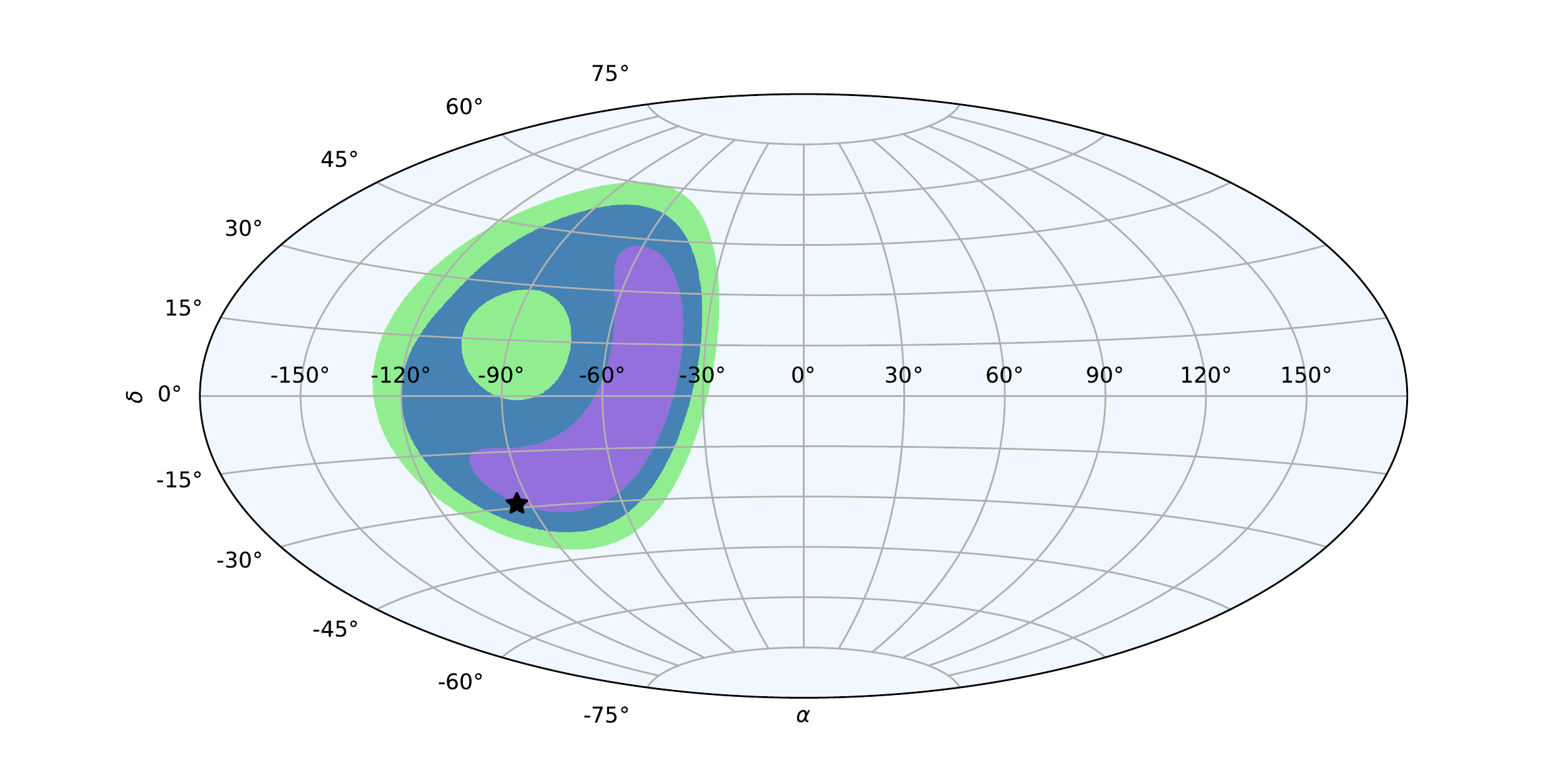}
\includegraphics[scale=0.3]{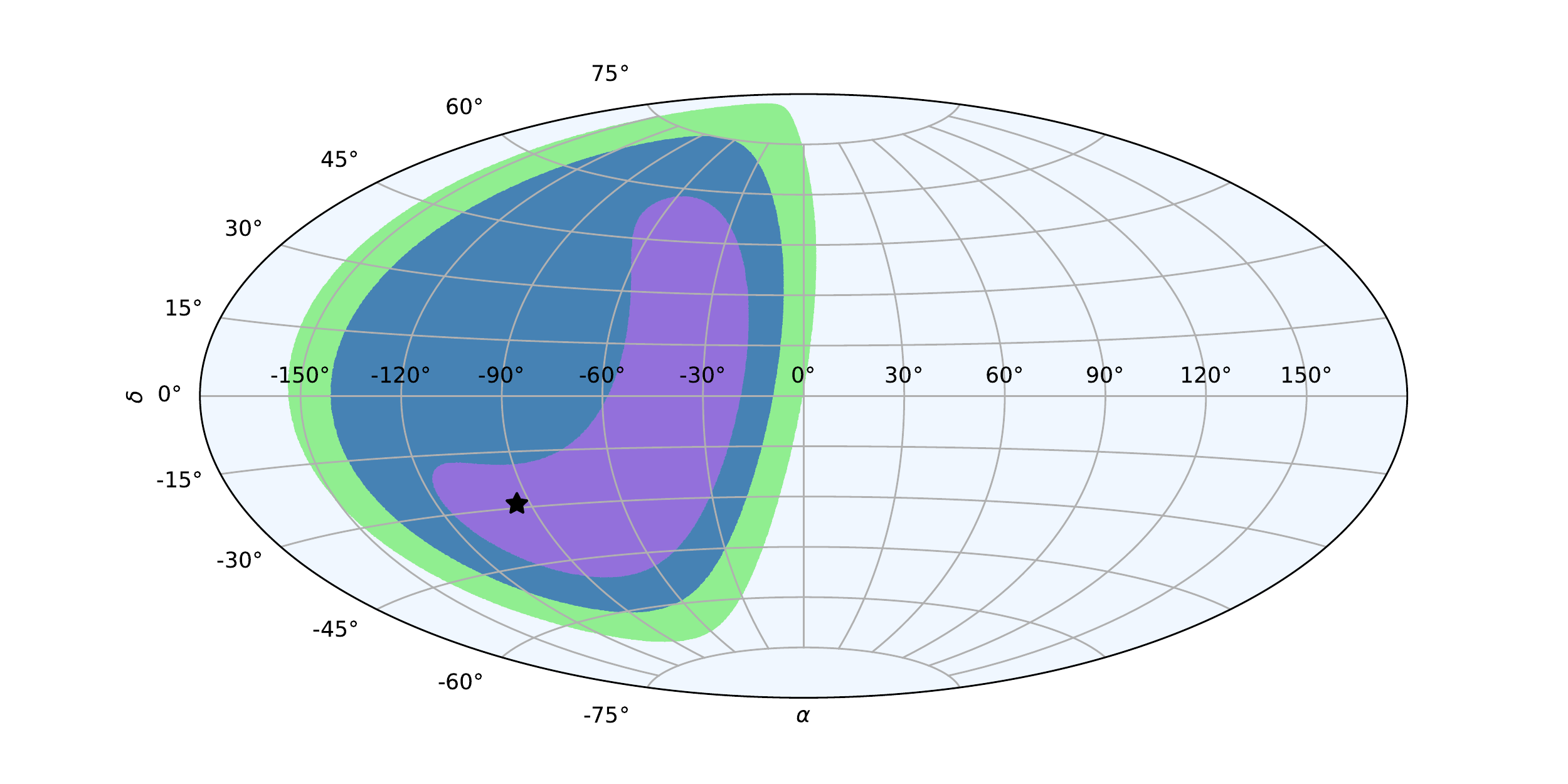}
\includegraphics[scale=0.3]{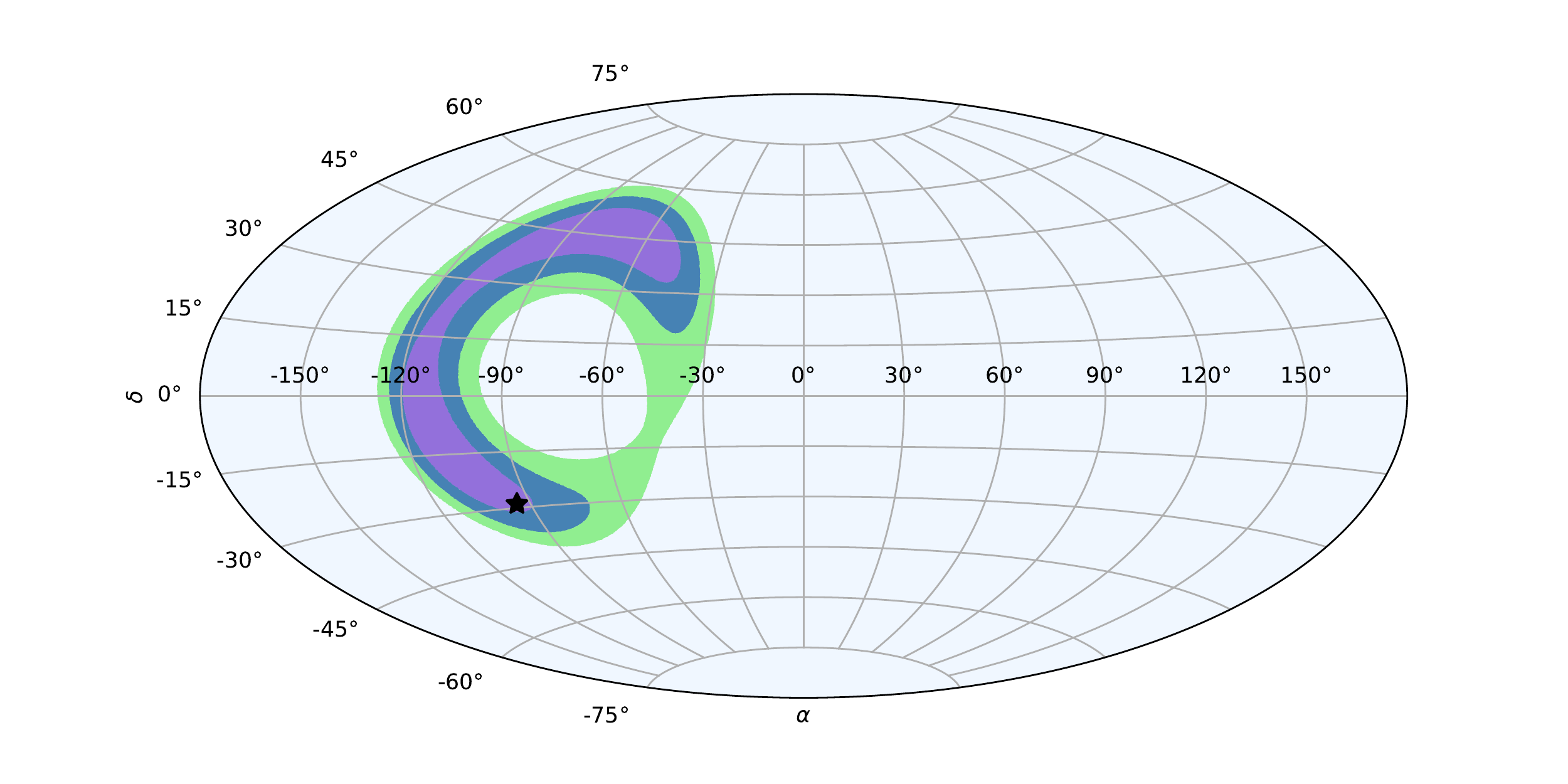}
\includegraphics[scale=0.3]{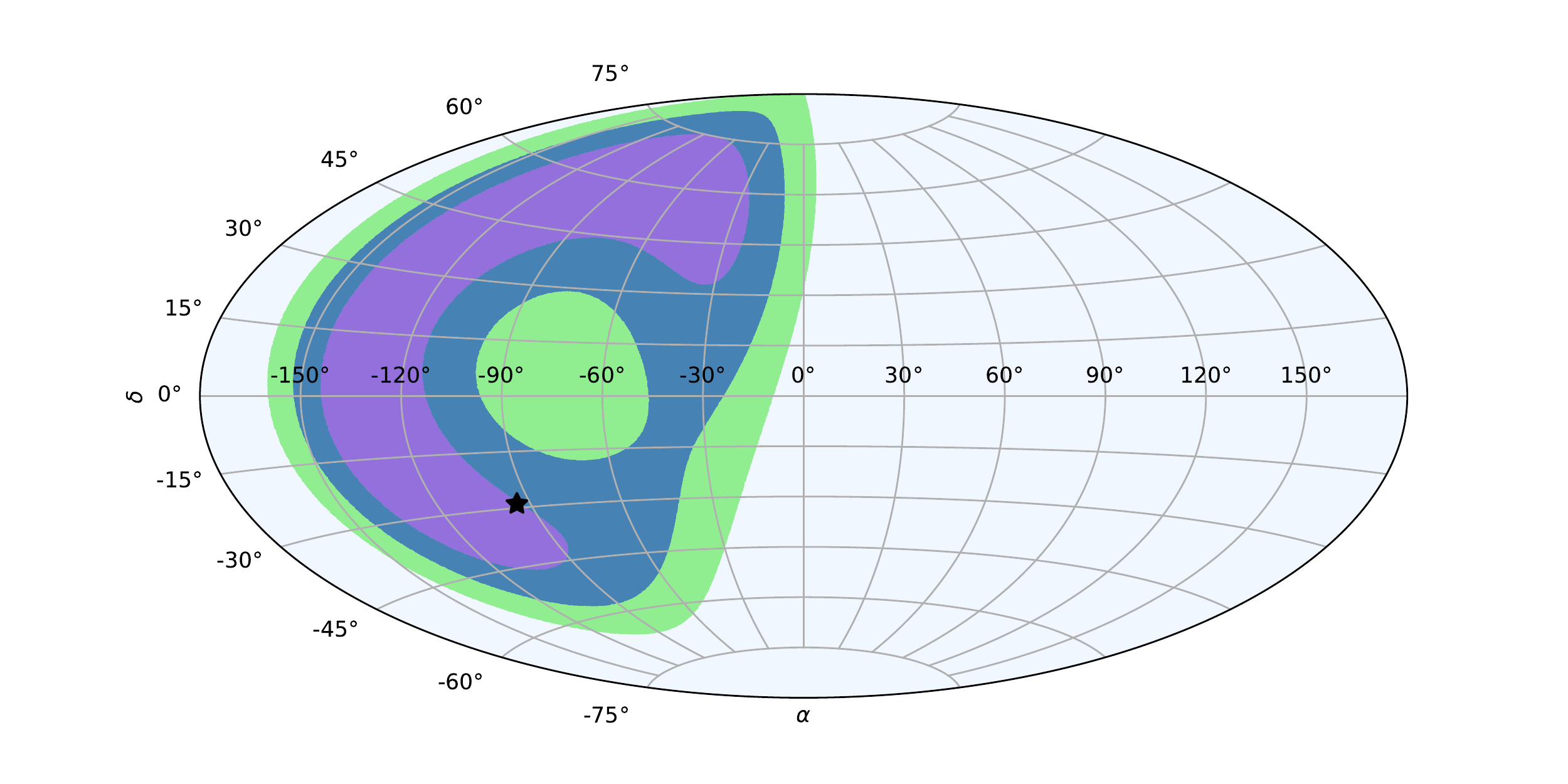}
\caption{1$\sigma$, 2$\sigma$, and 3$\sigma$ areas on the sky for a 10-kpc supernova. The plots use timing information from JUNO, DUNE, and Super-K (above) or Hyper-K (below) with IO (left) or NO (right), without the bias correction described in Sec.~\ref{sec:firstevent}. The true supernova location is marked with the black star.}
\label{fig:ih_nh_sk_hk_nocorrect}
\end{figure*}

\begin{figure*}[ht]
\centering
\includegraphics[scale=0.3]{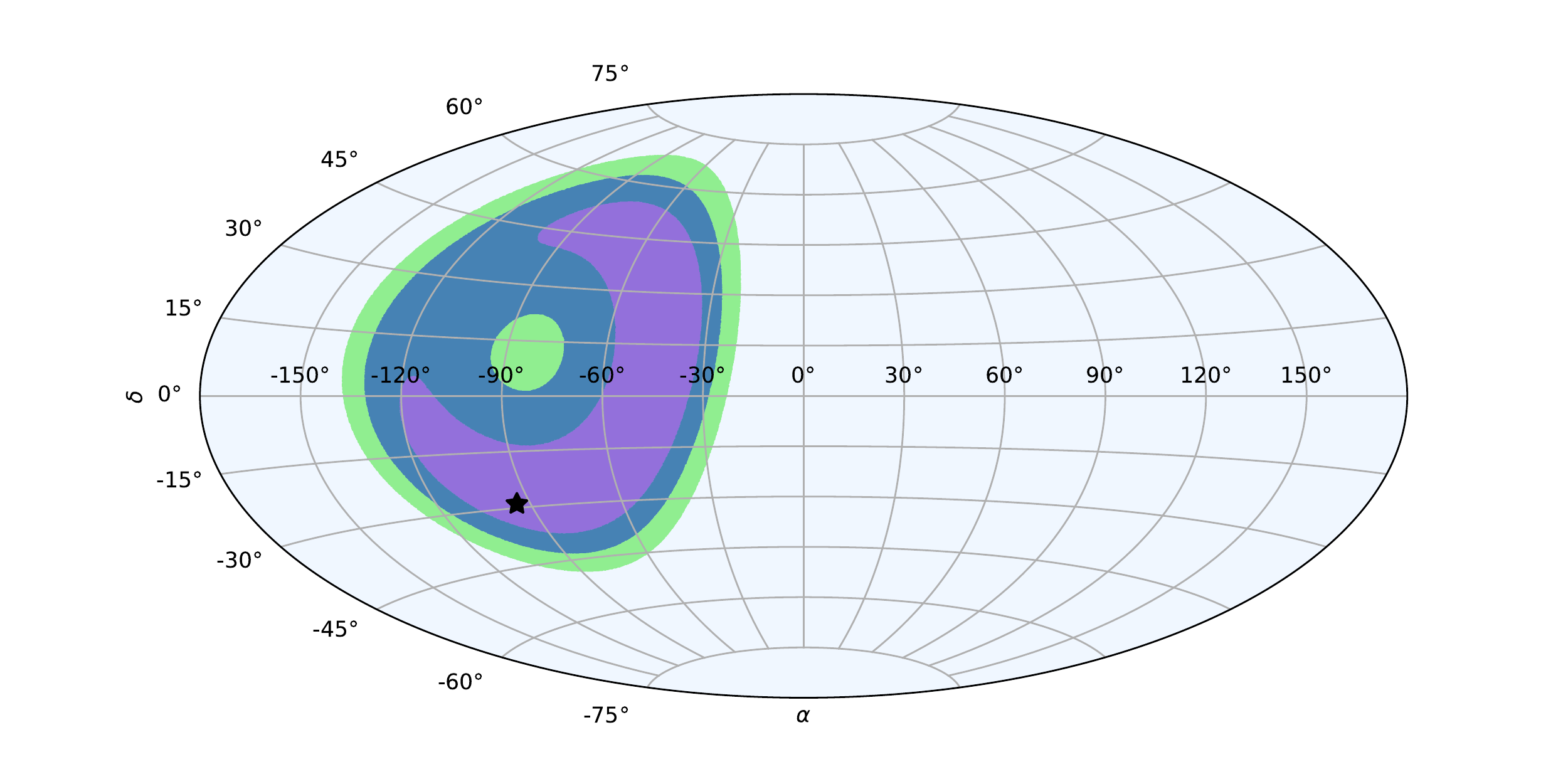}
\includegraphics[scale=0.3]{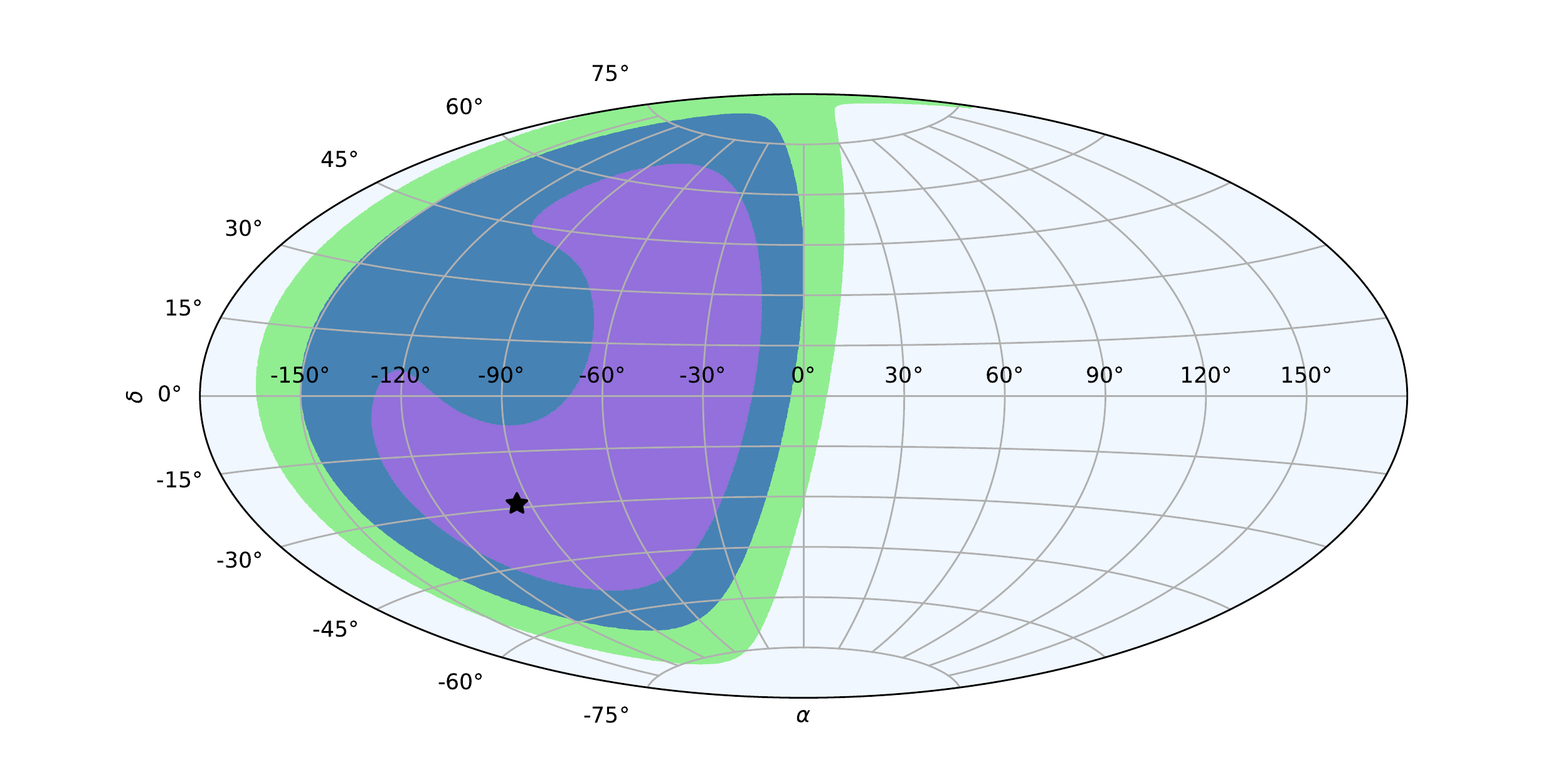}
\includegraphics[scale=0.3]{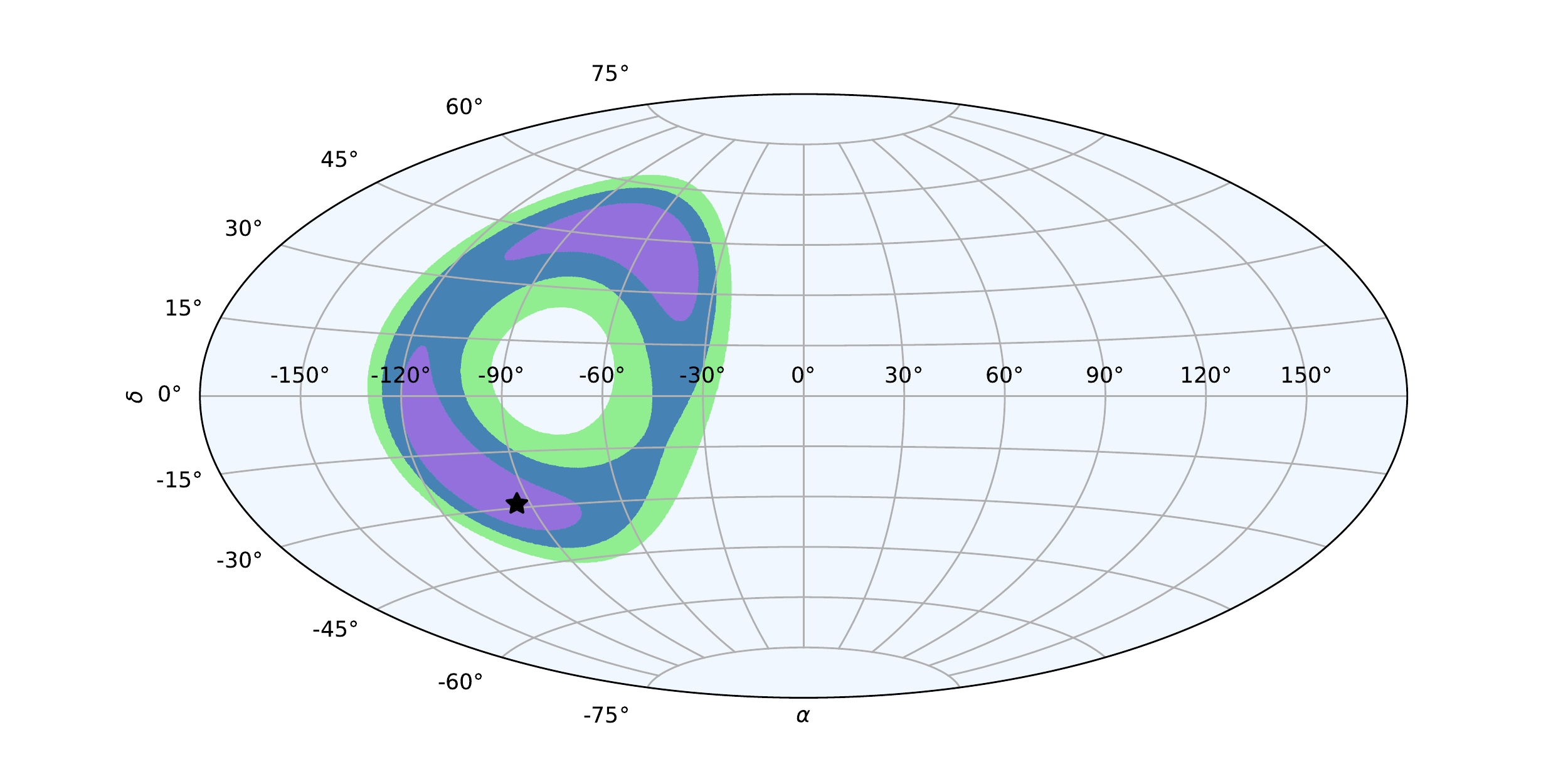}
\includegraphics[scale=0.3]{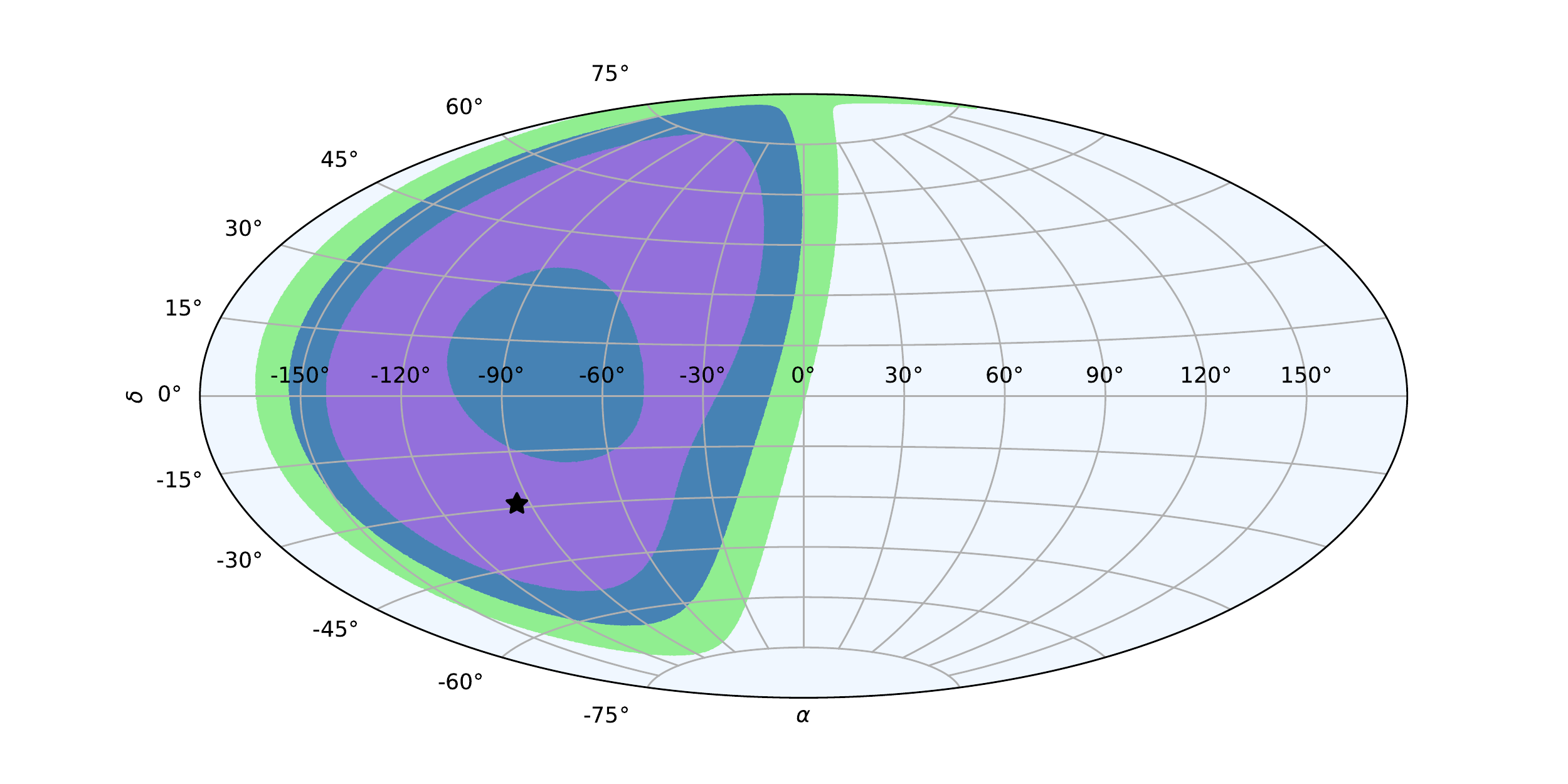}
\caption{1$\sigma$, 2$\sigma$, and 3$\sigma$ areas on the sky for a 10-kpc supernova. The plots use timing information from JUNO, DUNE, and Super-K (above) or Hyper-K (below) with IO (left) or NO (right).  These and subsequent plots include the bias correction. The true supernova location is marked with the black star.}
\label{fig:ih_nh_sk_hk}
\end{figure*}

We now examine results of the triangulation precision evaluation for various assumptions.  
Results for the detector combination of JUNO, DUNE, and either Super-K or Hyper-K are shown in Figs.~\ref{fig:ih_nh_sk_hk_nocorrect} (without bias correction) and ~\ref{fig:ih_nh_sk_hk} (including bias correction).  As expected, due to increased statistics, the search area for the supernova is smaller with the use of Hyper-K. 

\begin{figure*}[ht]
\centering
\includegraphics[scale=0.3]{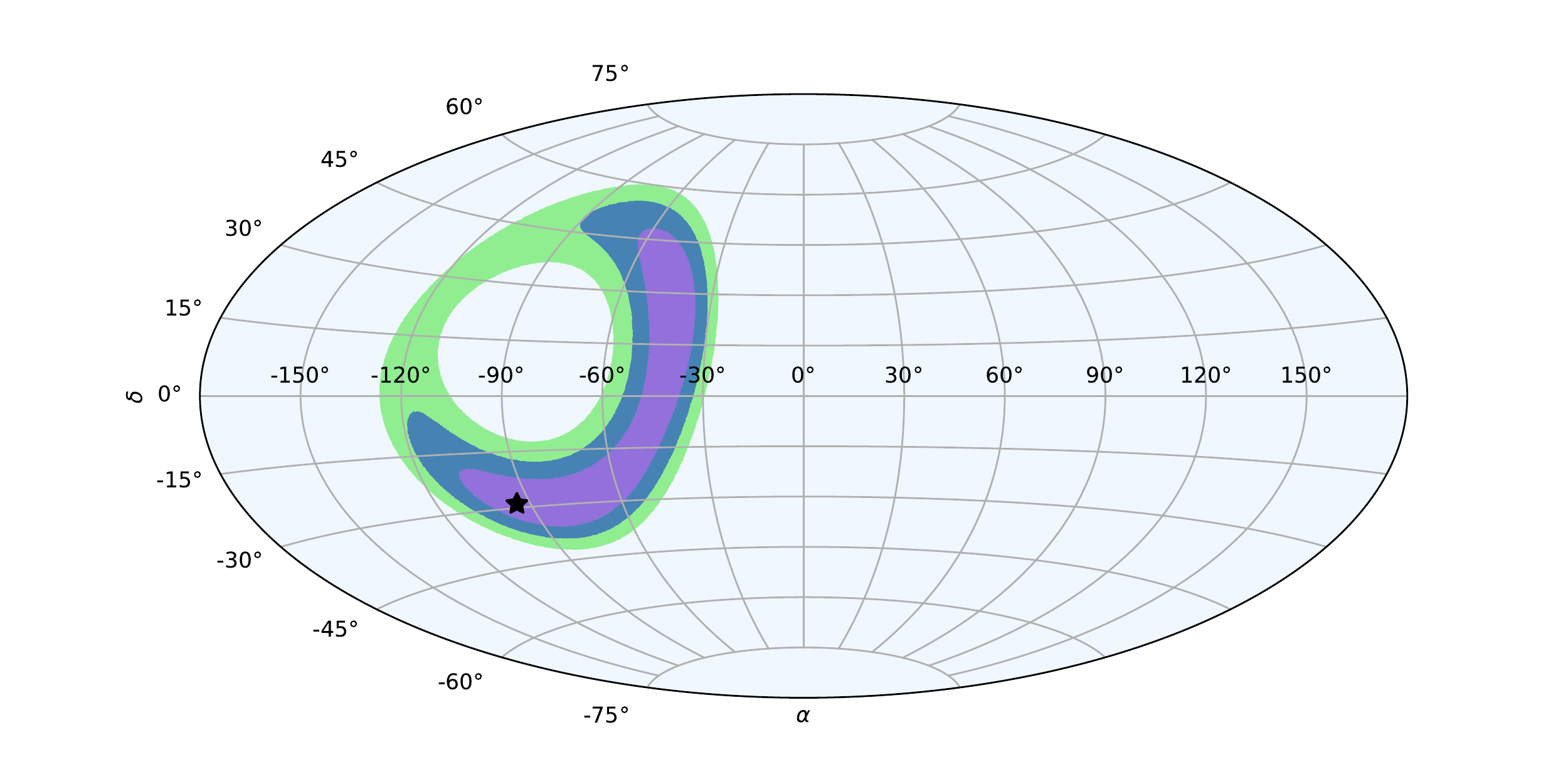}
\includegraphics[scale=0.3]{skymap/sk_dune_juno_ih_10.pdf}
\includegraphics[scale=0.3]{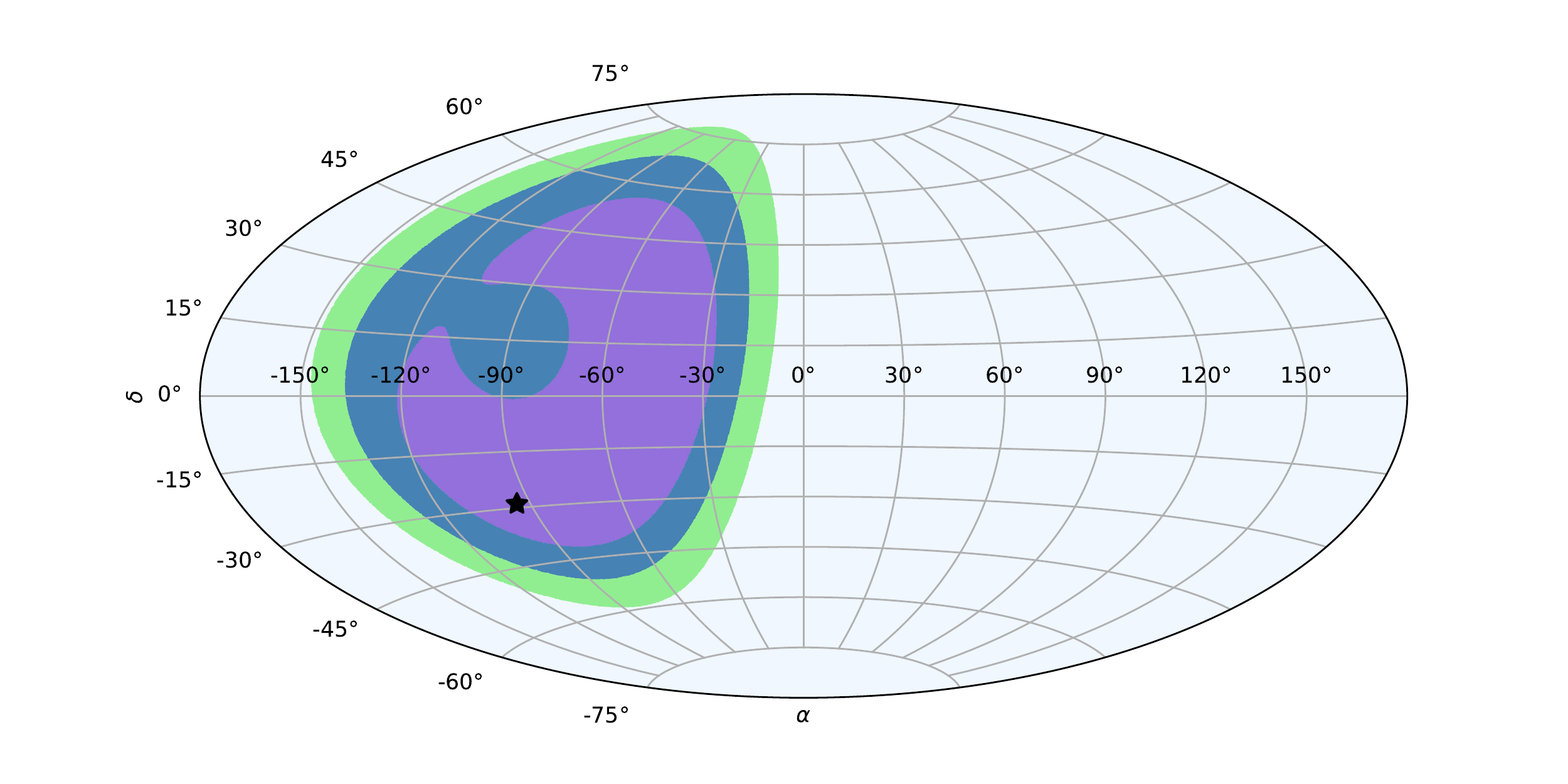}
\includegraphics[scale=0.3]{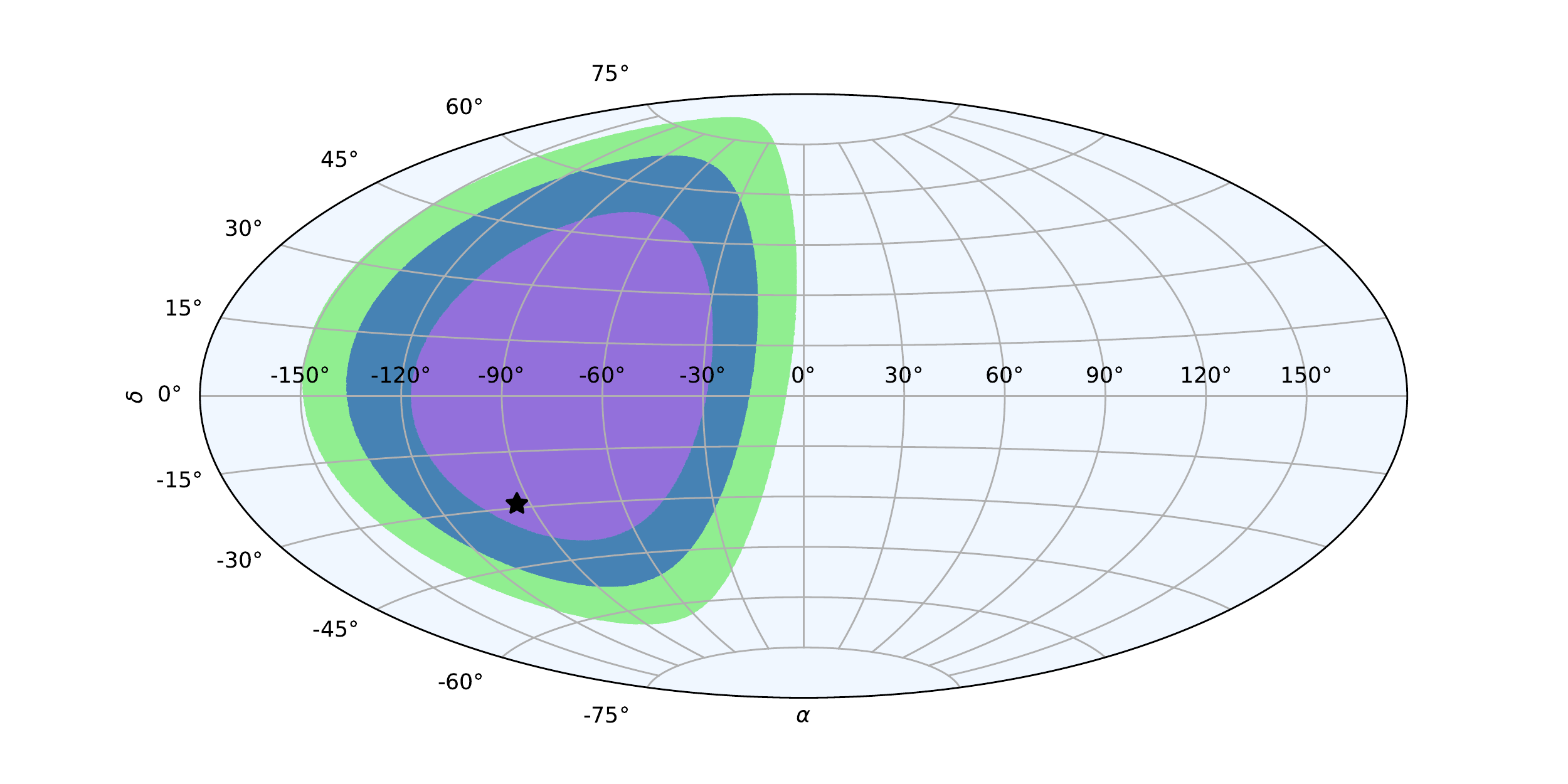}
\caption{Supernova search area as a function of distance from Earth. Top left: 5~kpc. Top right: 10~kpc.  Bottom left: 15~kpc.  Bottom right: 20~kpc. All calculations are made with DUNE, JUNO, and Super-K assuming IO.}
\label{fig:ih_sk_dist}
\end{figure*}

\begin{figure*}[ht]
\centering
\includegraphics[scale=0.3]{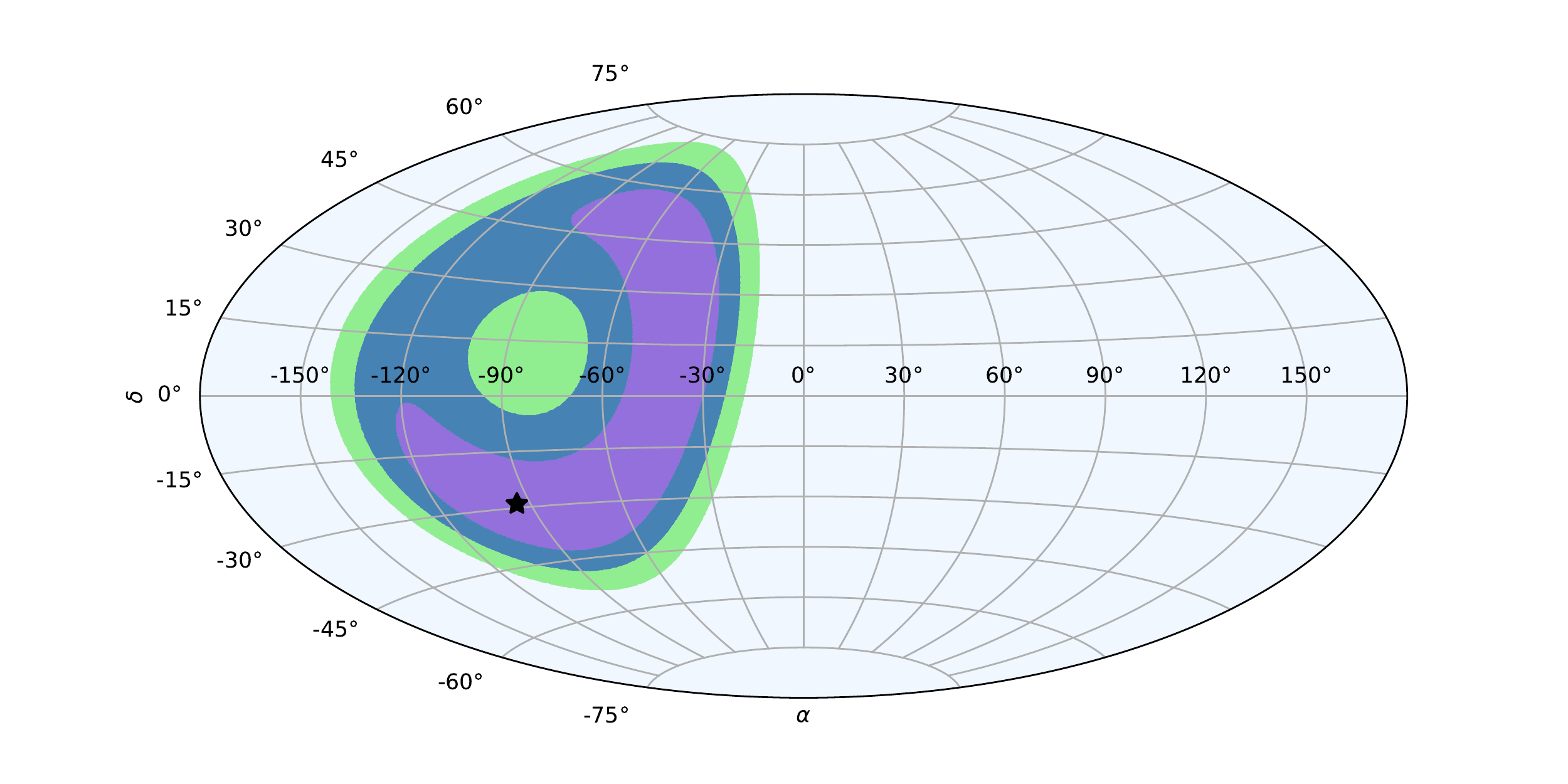}
\includegraphics[scale=0.3]{skymap/sk_dune_juno_nh_10.pdf}
\includegraphics[scale=0.3]{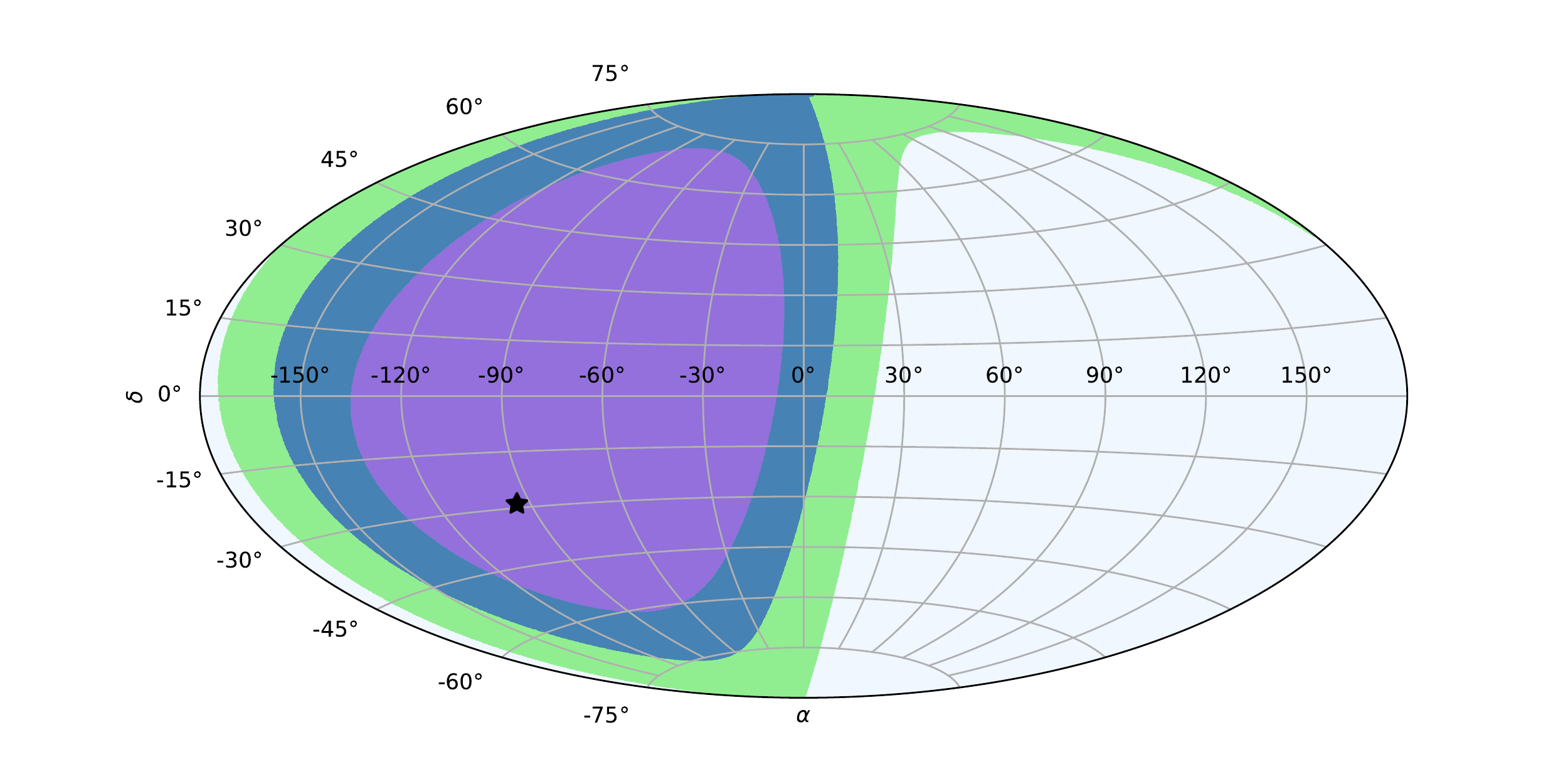}
\includegraphics[scale=0.3]{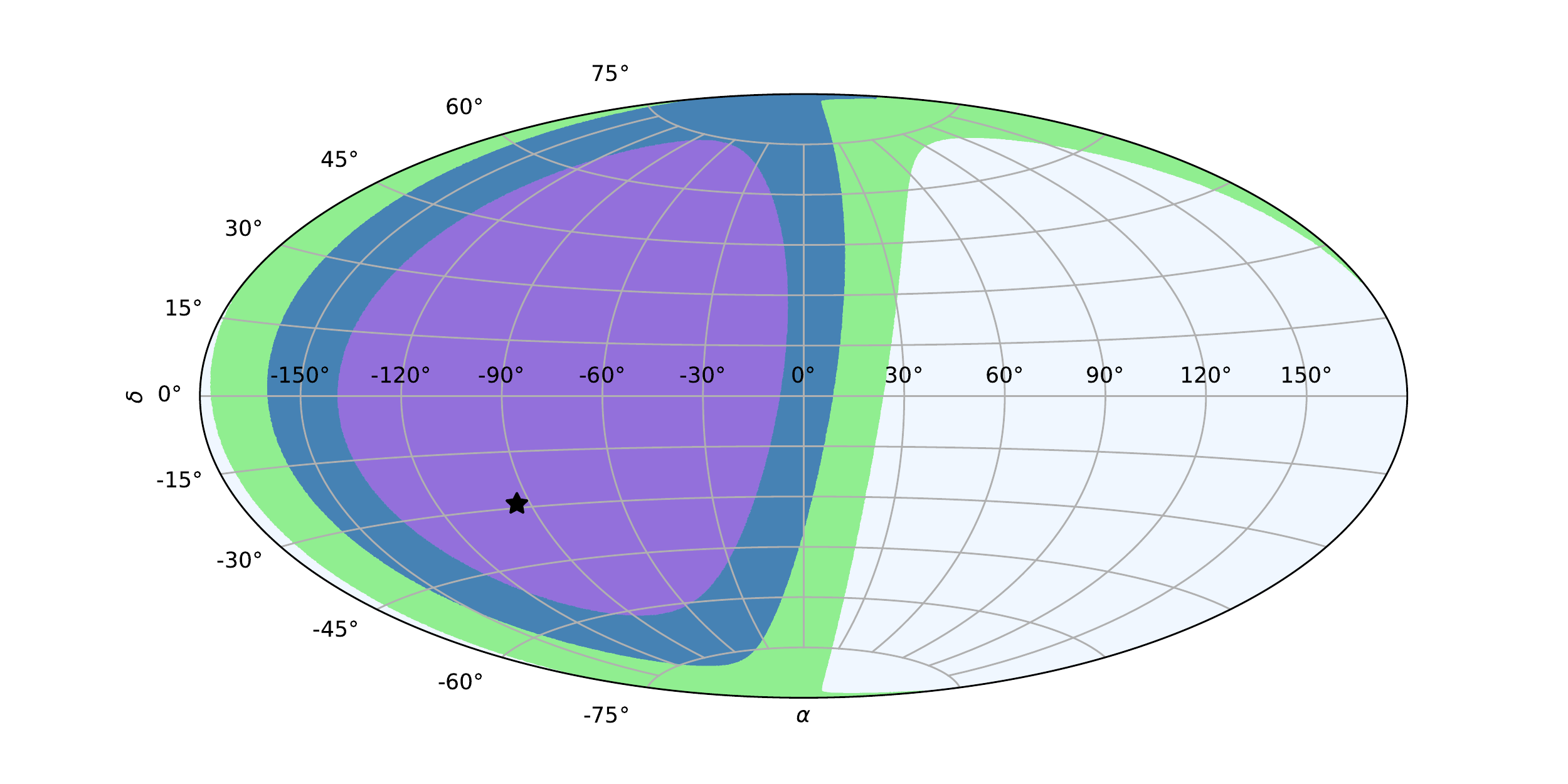}
\caption{Supernova search area as a function of distance from Earth. Top left: 5~kpc Top right: 10~kpc.  Bottom left: 15~kpc.  Bottom right: 20~kpc. All calculations are made with DUNE, JUNO, and Super-K assuming NO.}
\label{fig:nh_sk_dist}
\end{figure*}

As distance to the supernova increases, the number of neutrino events will decrease as inverse square of distance. This will lead to greater error, and a larger search area as in Figs. \ref{fig:ih_sk_dist} and \ref{fig:nh_sk_dist}. The areas are generated for DUNE, JUNO, and Super-K. As the distance of the supernova increases from 5 kpc to 20 kpc, the search area grows much larger. 

\begin{figure}[ht]
\centering
\includegraphics[scale=0.4]{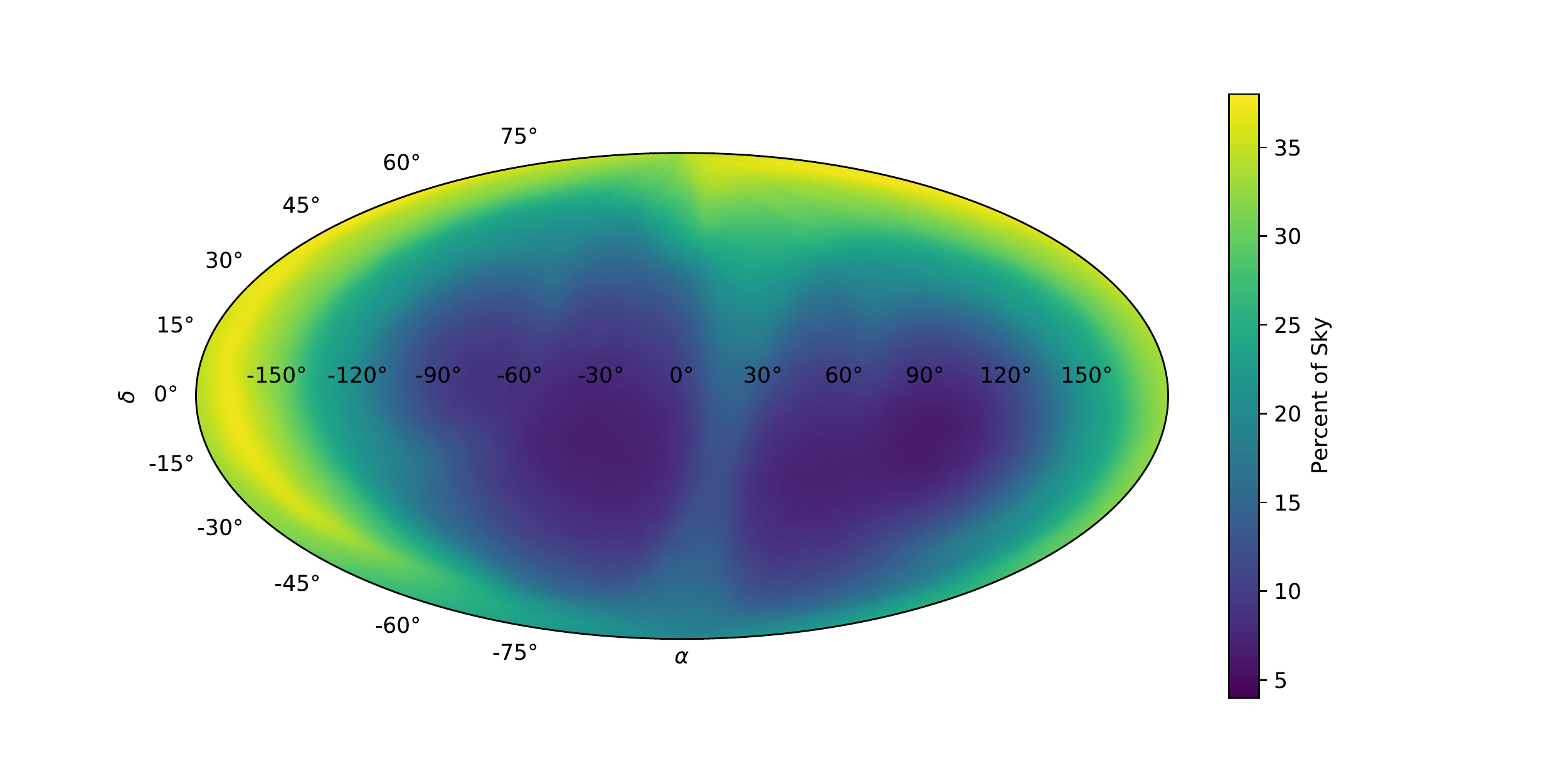}
\caption{1$\sigma$ sky fraction as a function of supernova location for the combination of JUNO, DUNE, and Super-K assuming NO, for a 10-kpc supernova.}
\label{fig:avg_sk_nh}
\end{figure}

\begin{figure}[ht]
\centering
\includegraphics[scale=0.4]{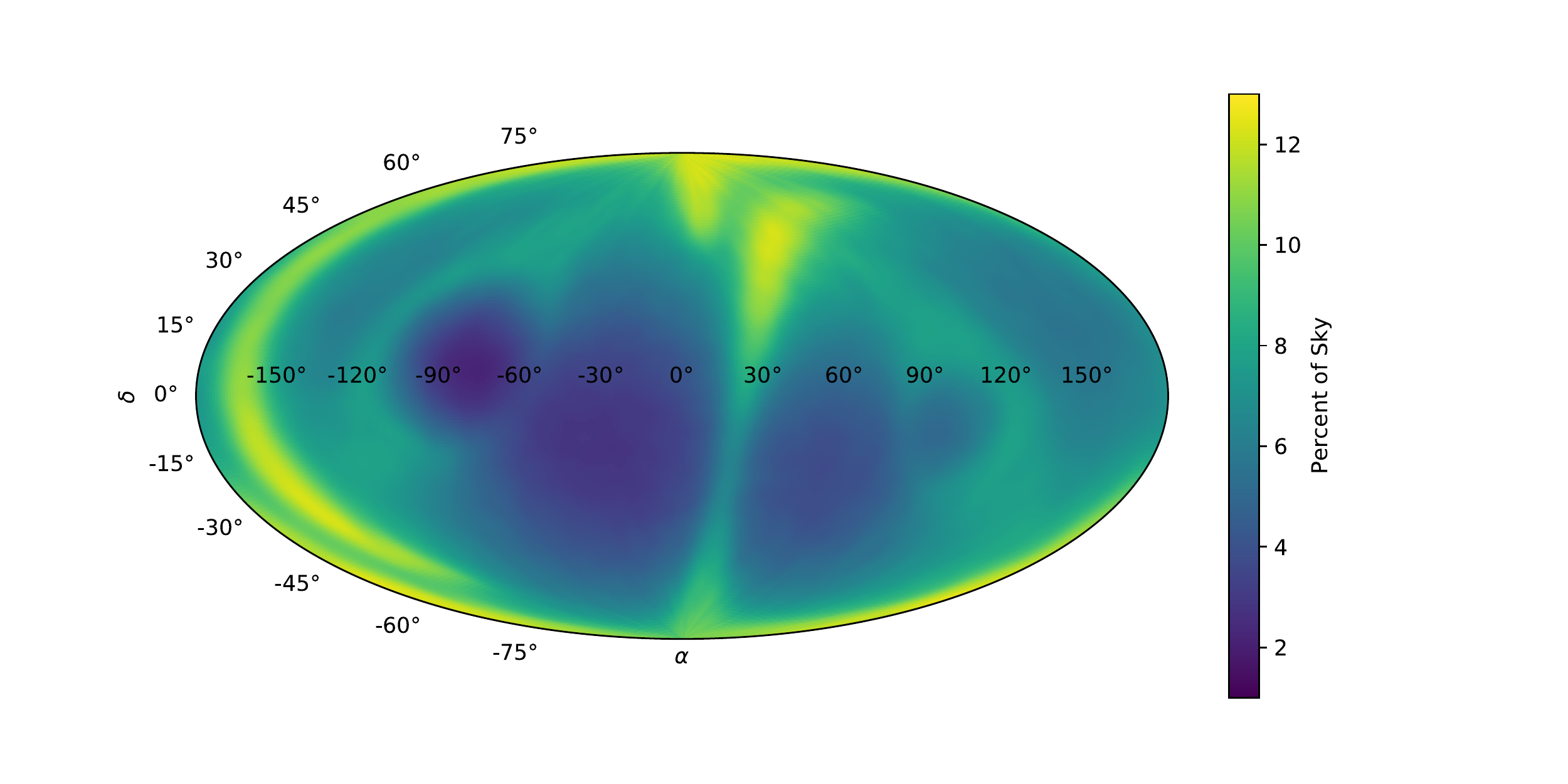}
\caption{1$\sigma$ sky fraction as a function of supernova location for the combination of JUNO, DUNE, and Super-K assuming IO, for a 10-kpc supernova.}
\label{fig:avg_sk_ih}
\end{figure}

\begin{figure}[ht]
\centering
\includegraphics[scale=0.4]{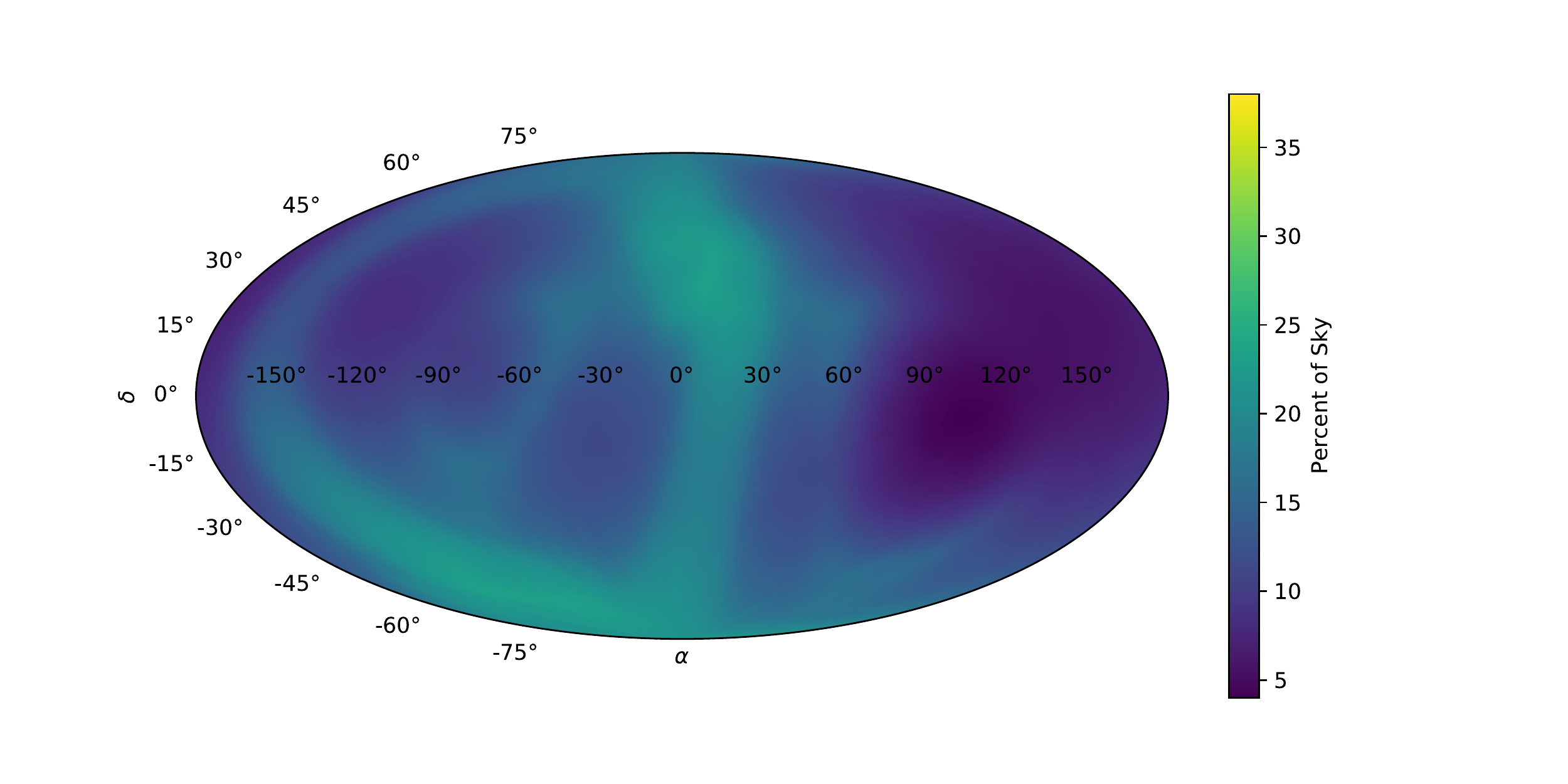}
\caption{1$\sigma$ sky fraction as a function of supernova location for the combination of JUNO, DUNE, and Hyper-K assuming NO, for a 10-kpc supernova.}
\label{fig:avg_hk_nh}
\end{figure}

\begin{figure}[ht]
\centering
\includegraphics[scale=0.4]{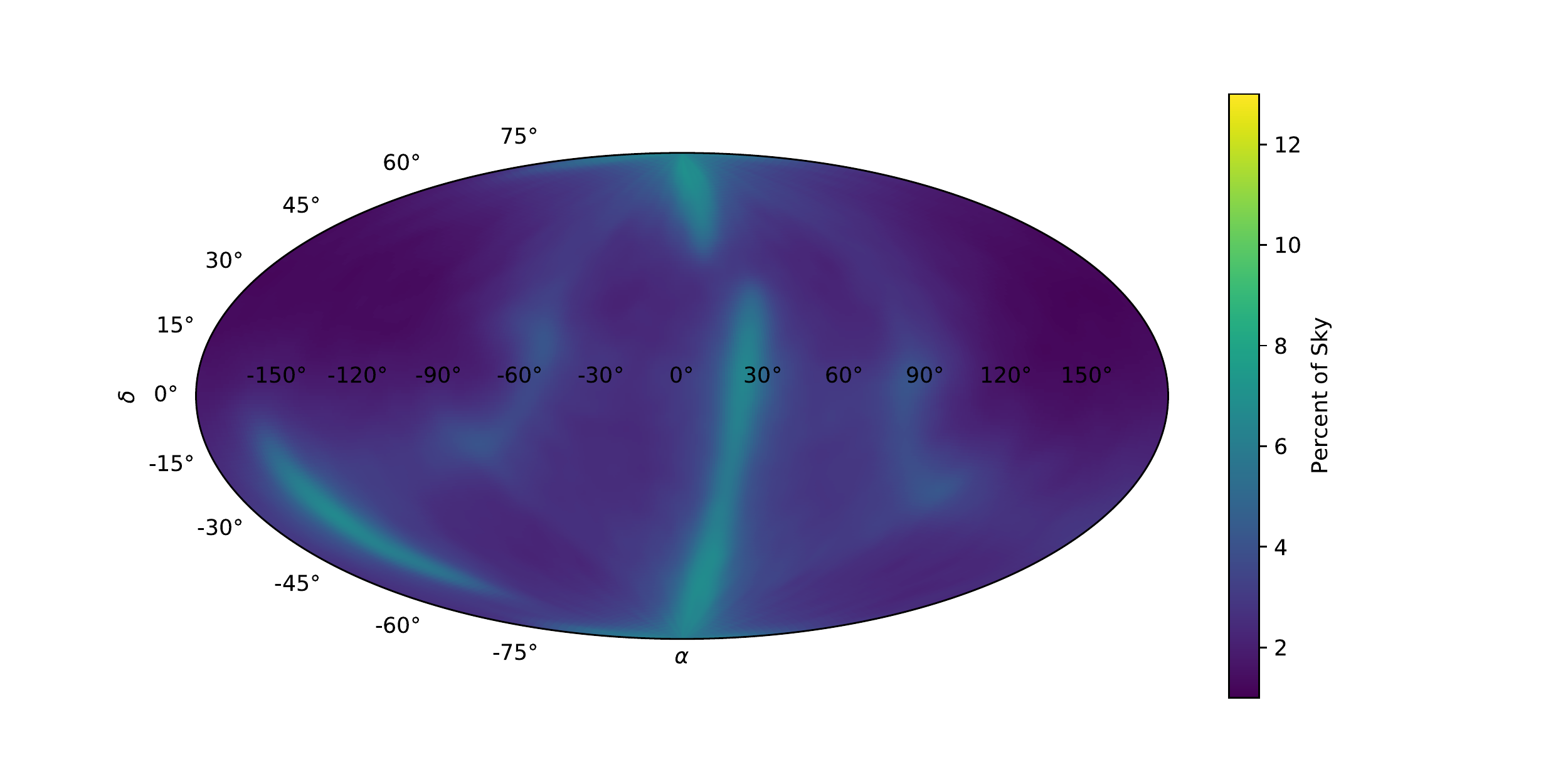}
\caption{1$\sigma$ sky fraction area as a function of supernova location for the combination of JUNO, DUNE, and Hyper-K assuming IO, for a 10-kpc supernova.}
\label{fig:avg_hk_ih}
\end{figure}

We then examine the effect of supernova sky location on the triangulation precision. In Figs. \ref{fig:avg_sk_nh} through \ref{fig:avg_hk_ih}, we see the percent of sky in the 1$\sigma$ area as a function of supernova location. All locations are sampled at one time--- the vernal equinox at noon. The structures in the plot are due to the specific relative positions on the globe of DUNE, JUNO, and Super-K or Hyper-K. In general, when Hyper-K is used, the area is smaller. The area is smaller when the time difference between detectors is largest, and larger when the time difference is smallest, as expected.

The addition of IceCube, as described in Sec.~\ref{sec:icecube}, results in a significant improvement in pointing precision.   This motivates further dedicated study of this possibility.

\begin{figure*}[ht]
\centering
\includegraphics[scale=0.3]{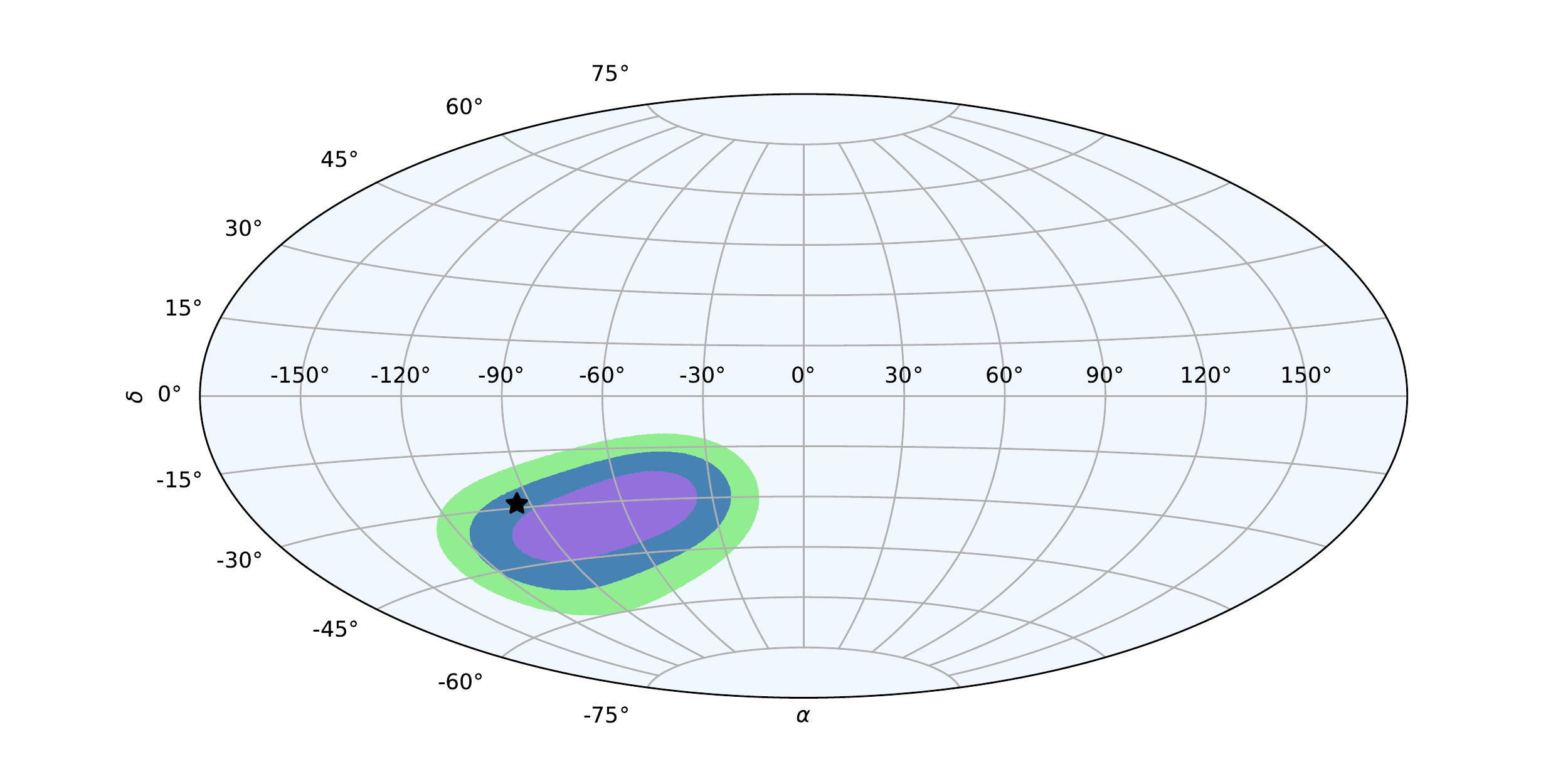}
\includegraphics[scale=0.3]{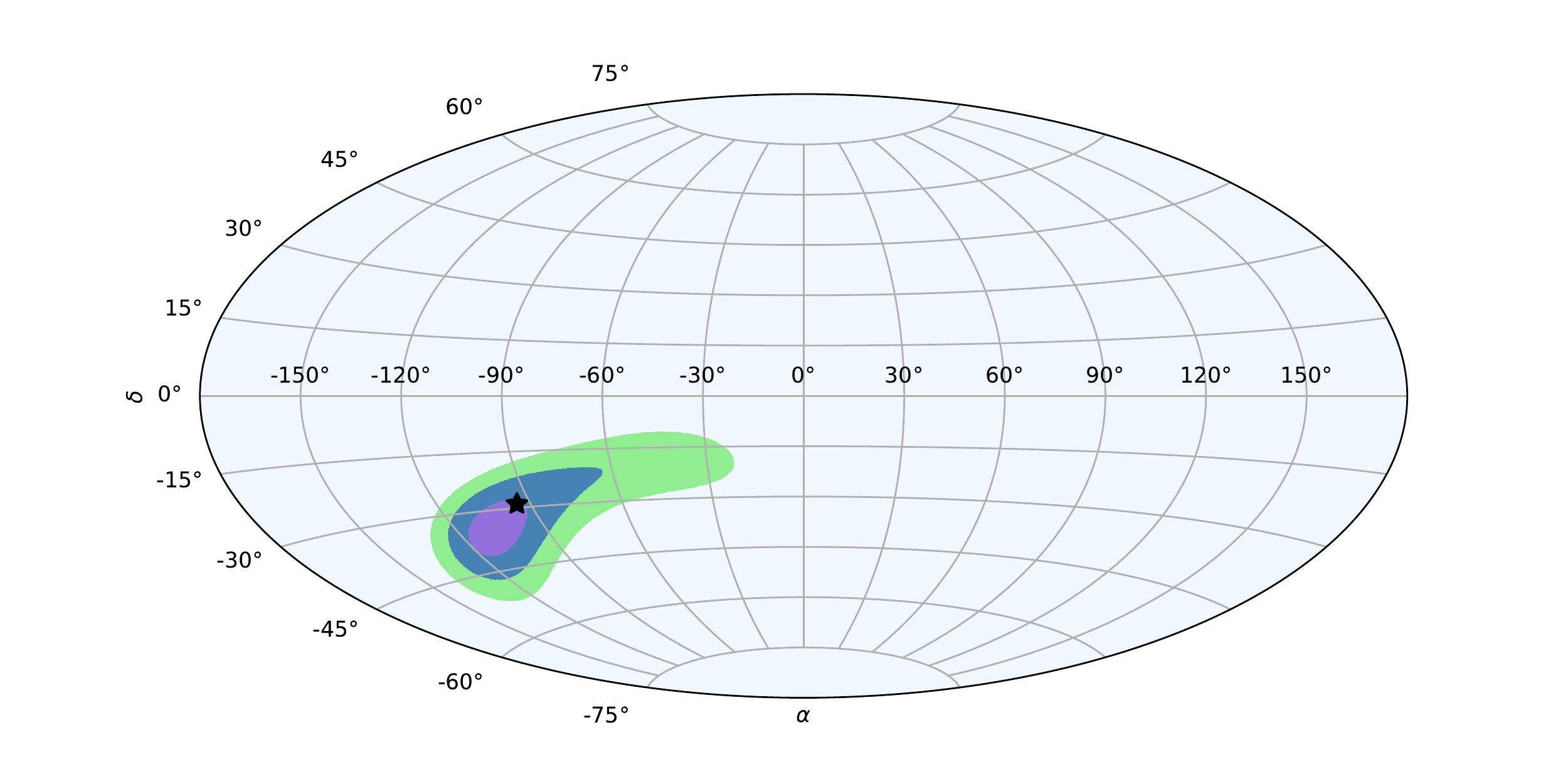}
\caption{Sky areas determined by combining IceCube timing information with DUNE, JUNO, and Super-K (left) or Hyper-K (right) at 10~kpc and assuming NO.}
\label{fig:nh_ic}
\end{figure*}

\begin{figure}[ht]
\centering
\includegraphics[scale=0.3]{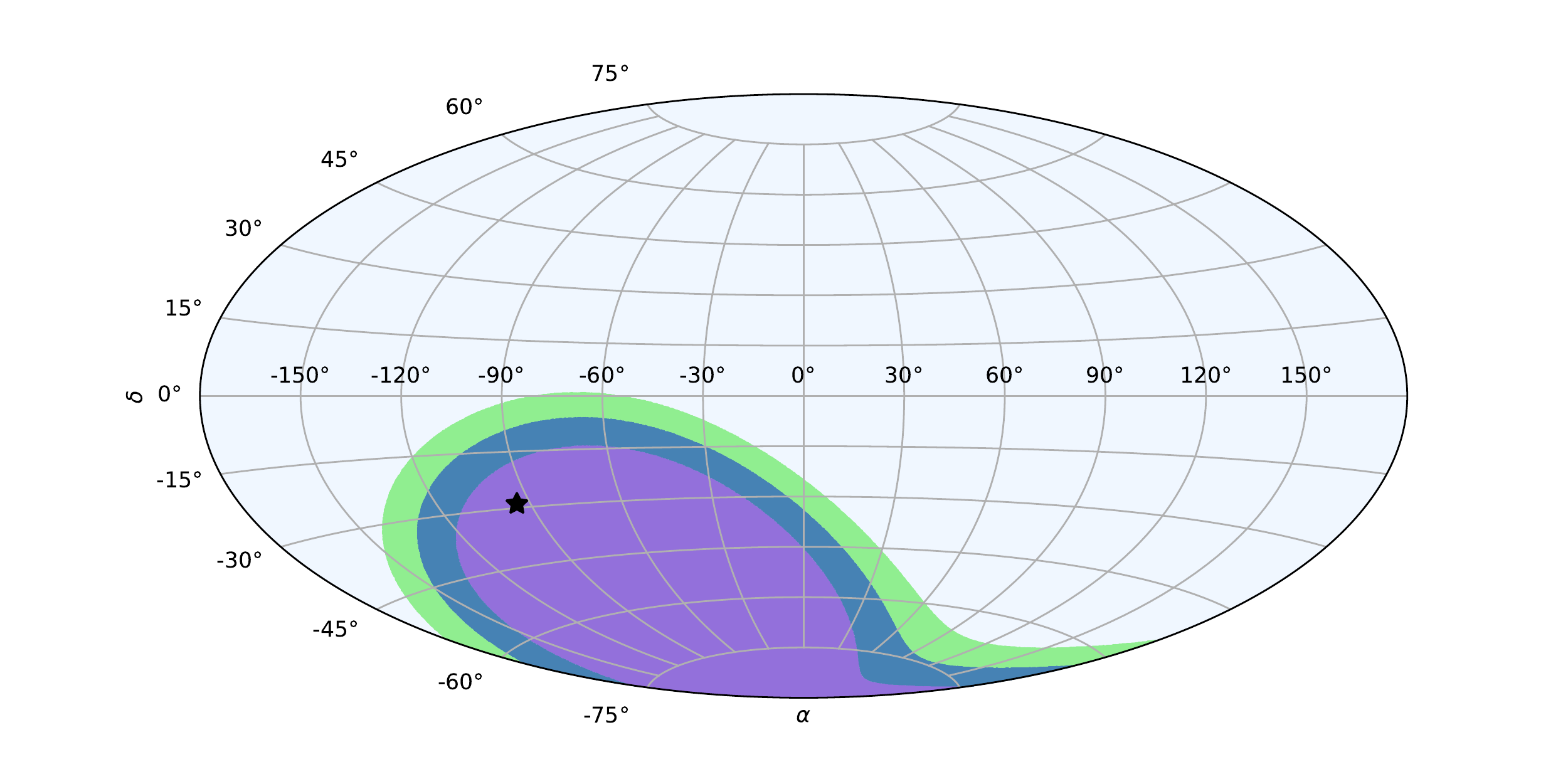}
\caption{Sky areas determined by combining timing information from JUNO and IceCube at 10~kpc and assuming NO.  This information will be available even if there is no anisotropic interaction information available from detectors with intrinsic directionality.}
\label{fig:ic_juno}
\end{figure}

For a case when the Cherenkov detectors and DUNE (which have high-quality individual-detector pointing thanks to intrinsic directionality) are offline, the combination of JUNO and IceCube may have the best pointing capabilities. With only these detectors, the search area is still substantially reduced from the whole sky, as seen in Fig.~\ref{fig:ic_juno}.

\begin{figure}[ht]
\centering
\includegraphics[scale=0.3]{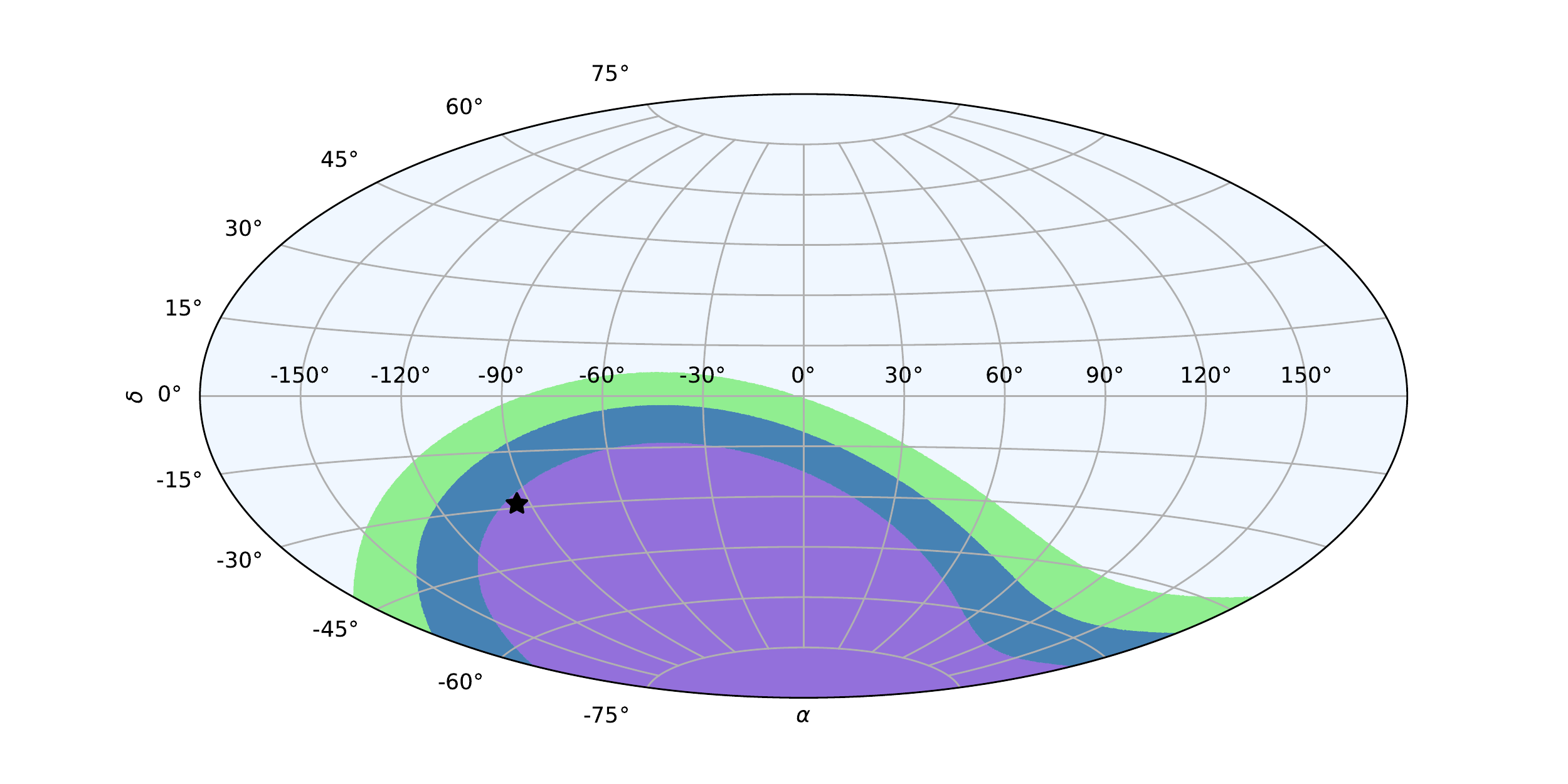}
\caption{Sky areas determined by combining timing information from SK and IceCube at 10~kpc and assuming NO.  This represents information from detectors which are currently active.}
\label{fig:sk_ic}
\end{figure}

Of the five neutrino detectors considered, only Super-K and IceCube are currently online. The result of combining timing information from only these detectors is shown in Fig.~\ref{fig:sk_ic}, and represents current triangulation pointing capabilities.

\begin{figure}[ht]
\centering
\includegraphics[scale=0.3]{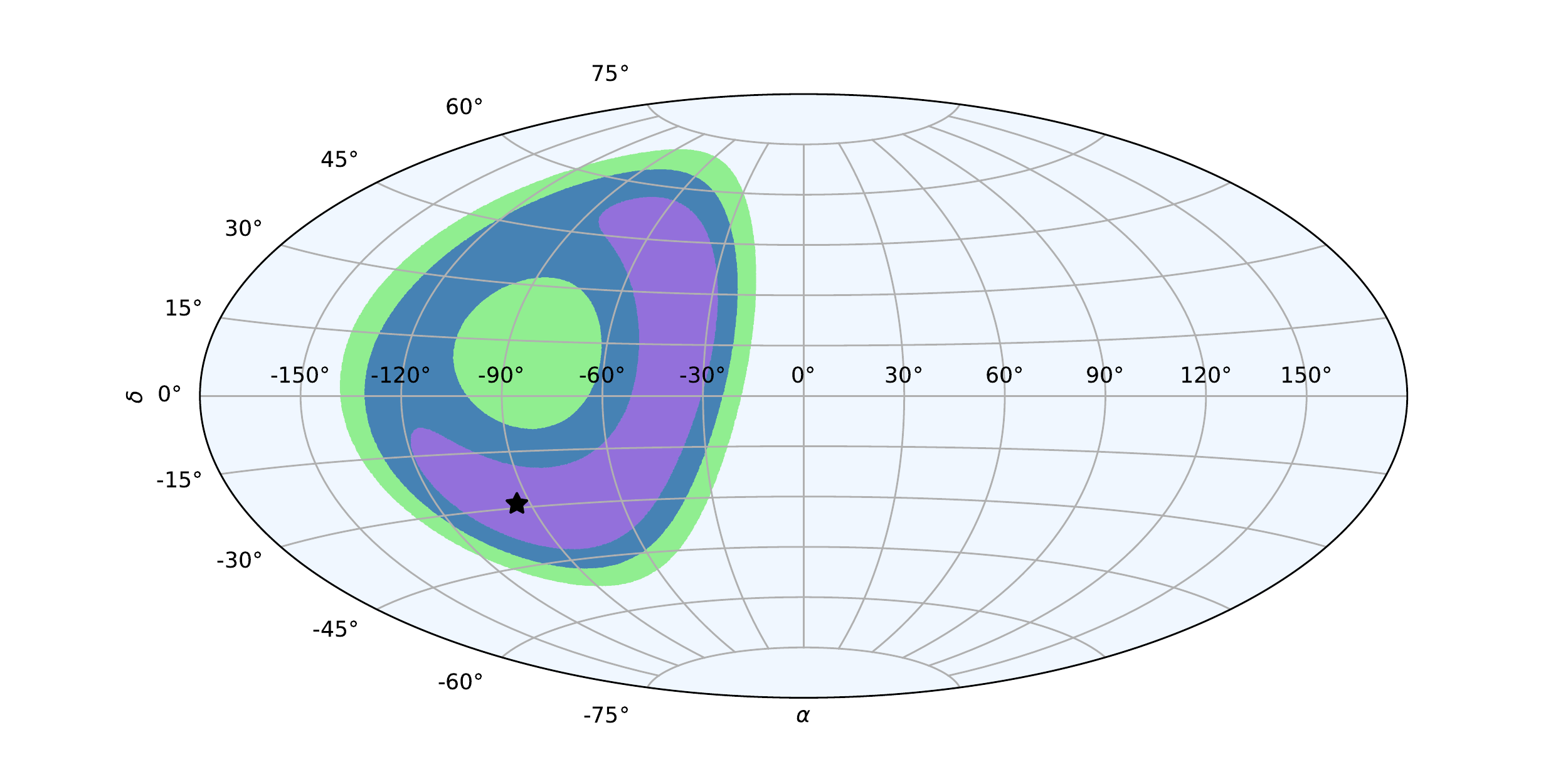}
\includegraphics[scale=0.3]{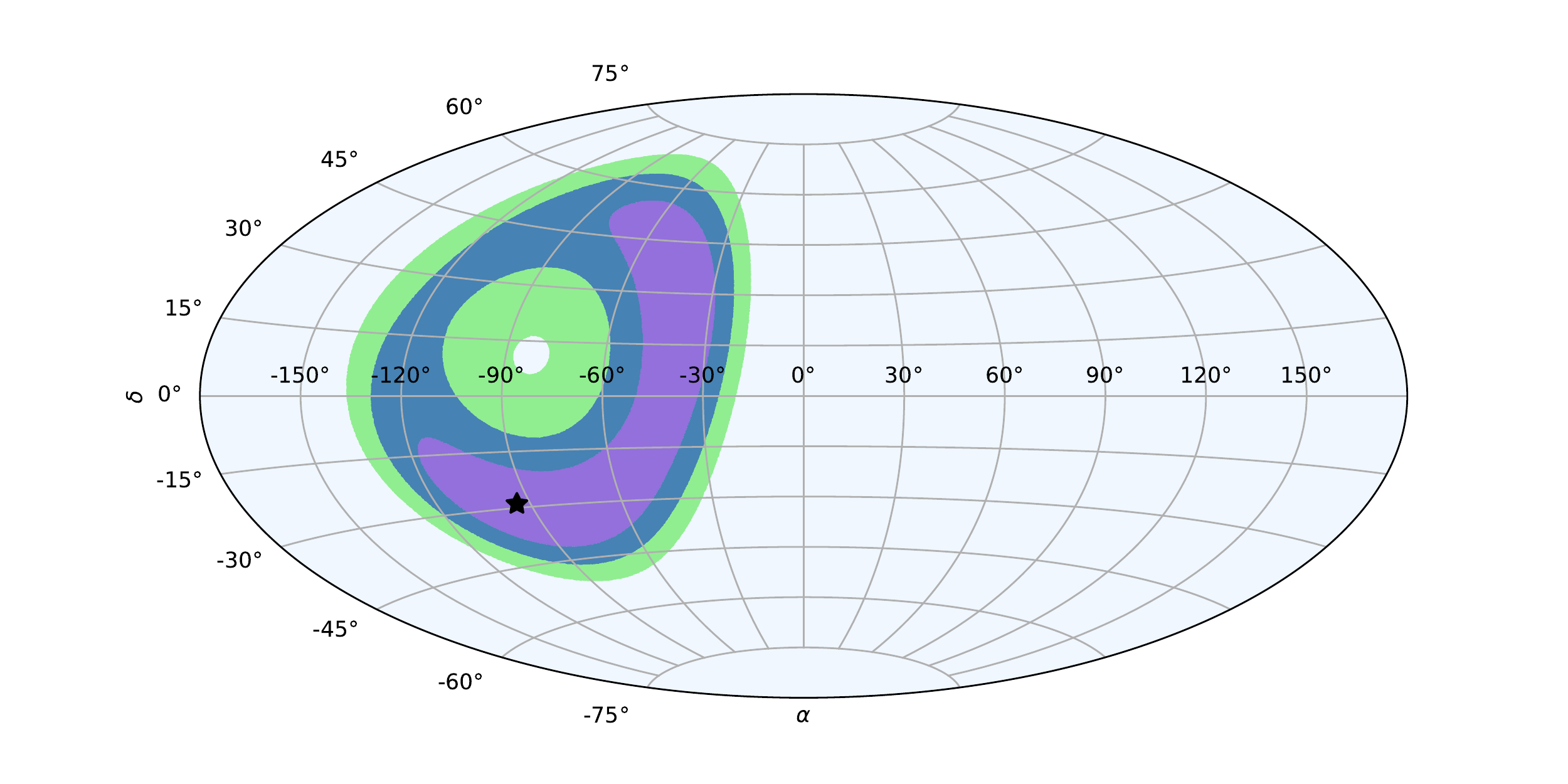}
\caption{Areas corresponding to 1$\sigma$, 2$\sigma$, and 3$\sigma$ at 10~kpc generated using time differences between DUNE, JUNO, and Super-K with NO and models from H\"udepohl for different progenitor masses. Left: 11.2~M\textsubscript{\(\odot\)}. Right: 27.0~M\textsubscript{\(\odot\)}.}
\label{fig:hudepohl}
\end{figure}

\begin{figure}[ht]
\centering
\includegraphics[scale=0.3]{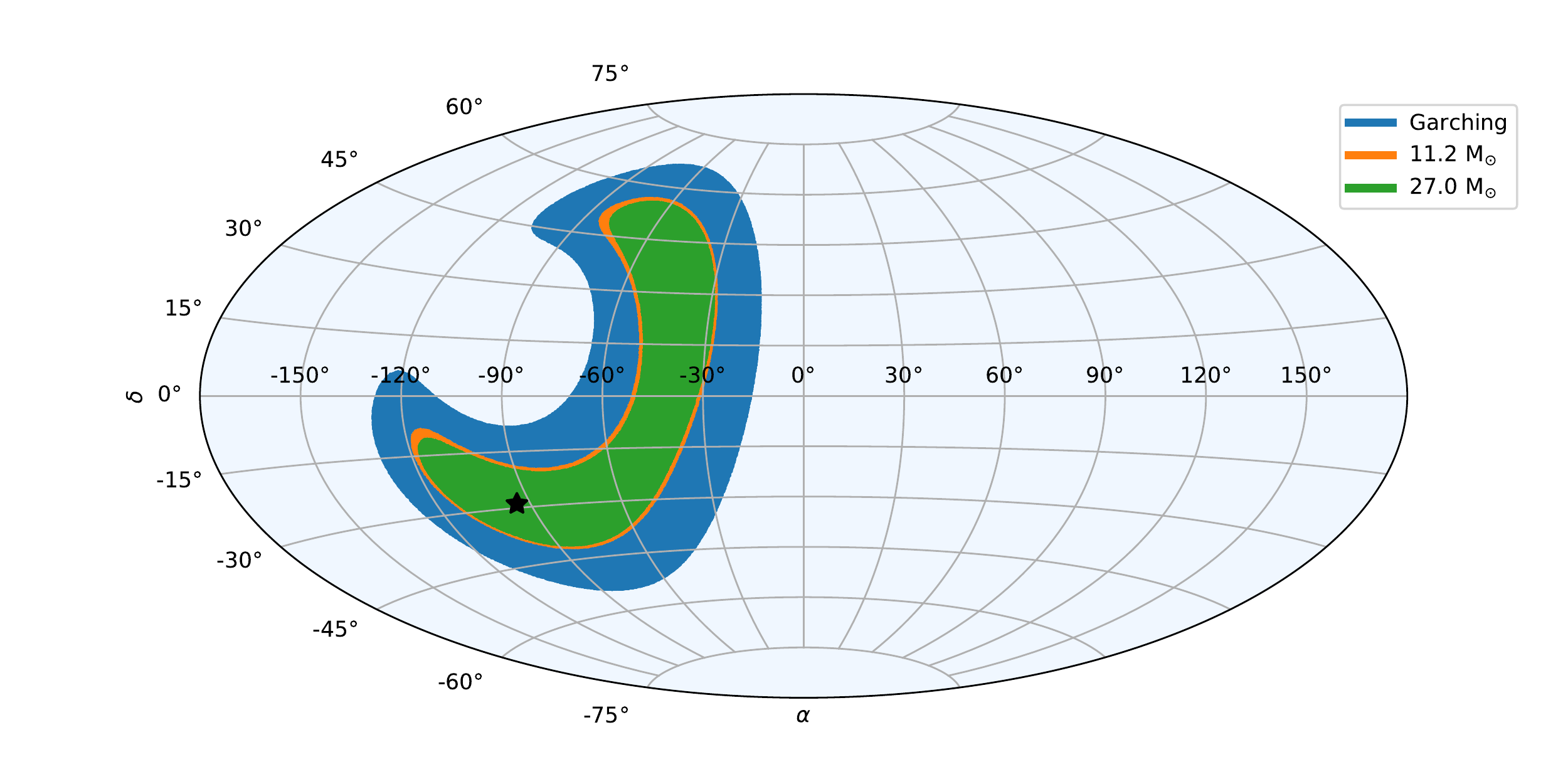}
\caption{Areas corresponding to 1$\sigma$ at 10~kpc generated using time differences between DUNE, JUNO, and Super-K with NO for the Garching model and models from H\"udepohl for two different progenitor masses.}
\label{fig:models}
\end{figure}

Additionally, we explore a few other supernova models.  We have considered those described in Ref.~\cite{Hudepohl2013} with equation of state from Ref.~\cite{Shen1998}. We look at core-collapse supernovae with masses of 11.2~M\textsubscript{\(\odot\)} and 27.0 M\textsubscript{\(\odot\)} (Cooling-Shen\_s11.2 and Cooling-Shen\_s27.0), and detections with DUNE, JUNO, and Super-K with the NO assumption. Sky areas are shown in Fig.~\ref{fig:hudepohl}. In both cases, the area included within 1$\sigma$ error is smaller than that when using the Garching model, as shown in Fig.~\ref{fig:models}.  This is primarily due to the increased neutrino event rate at the start of the neutrino burst for these models.

Our results are overall less optimistic than those in Ref.~\cite{Brdar2018}.  We checked that our evaluated variances on the time differences with respect to the true neutrino signal for individual detectors are quite similar (within less than 1 ms) to the results in that reference.
The main reason for our less optimistic results is that we are directly evaluating spread of time differences between detectors, which takes into account different detector response, rather than using the maximum $\sigma_t$ of individual-detector time differences with respect to an unknown true flux in the denominator of the $\chi^2_{ij}$.   Our results correspond to what will be possible to do in practice, and are furthermore fairly robust against supernova model choice.

\section{Conclusions}\label{sec:conclusion}

Under favorable conditions, and by combining timing information from many neutrino detectors, the direction to a core-collapse supernova can be triangulated using the relative event timing of neutrino signals observed in detectors around the world.  A simple first-event method is robust, and should be possible to apply promptly in practice.  A detector-dependent relative-timing bias can in principle be corrected for using the data themselves to improve the accuracy of the measurement.  A future study will consider practical ways of fast information sharing to optimize triangulation pointing.

The pointing precision is dependent on the location of the supernova, its distance from Earth, the specific detectors used, and their efficiencies and backgrounds. Additionally, neutrino mass ordering will affect the event rate in detectors via flavor transition differences, and therefore will affect the precision of the triangulation method.  
The triangulation method is not the only way to point to a core-collapse event using the neutrino burst information; in particular, elastic scattering of neutrinos on electrons in detectors with directional capability is likely to do significantly better.  However, there remains the possibility that a detector with such capability will not be online or may not be able to provide pointing information promptly.  Furthermore, optimal pointing information may well be obtained by combining information from different methods; the intersection of a triangulation band and an elastic-scattering spot may be better than either alone. 
It is also worth considering different practical strategies for different timescales.  Lower-precision triangulation pointing can be done fast, and then improved with successive refinements at later times by incorporating additional information as it becomes available.
Discussion of the possible overall global supernova neutrino pointing precision from use of all information available is beyond the scope of this work.  Realistic real-time strategies for optimization of worldwide neutrino pointing capabilities, considering improved knowledge of detector responses and backgrounds, will be a topic of future investigation.

\section*{Acknowledgments}

NBL was supported for summer work at Duke University by the Caltech Summer Undergraduate Research Fellowships program and the National Science Foundation.  The research activities of KS are supported by the Department of Energy and the National Science Foundation.  We are grateful to R.~Patterson and to Duke Neutrino and Cosmology Group members, especially D.~Pershey, E.~Conley and A.J.~Roeth.

\bibliography{references}
\end{document}